\documentclass[journal]{IEEEtran}

\usepackage{cite}
\usepackage{amsmath,amssymb,amsfonts}
\usepackage{algorithmic}
\usepackage{graphicx}
\usepackage{textcomp}

\usepackage{tikz} 
\usetikzlibrary{calc, 
                arrows.meta, 
                intersections, 
                fadings,                   
                decorations.pathreplacing, 
                positioning                
                }
\usepackage{tikz-3dplot} 
\usetikzlibrary{perspective} 
\usepackage{hyperref}
\hypersetup{
    colorlinks=true,   
    linkcolor=blue,    
    filecolor=magenta, 
    citecolor=blue,    
    urlcolor=blue      
}
\usepackage{orcidlink} 
\usepackage[nolist]{acronym} 
\usepackage{balance} 
\usepackage{esint} 
\usepackage[nameinlink]{cleveref} 
\usepackage{bm} 
\usepackage{units} 
\usepackage[caption=false, font=footnotesize]{subfig} 
\usepackage{svg}

\def\BibTeX{{\rm B\kern-.05em{\sc i\kern-.025em b}\kern-.08em
    T\kern-.1667em\lower.7ex\hbox{E}\kern-.125emX}}

\crefname{figure}{Fig.}{Fig.}
\crefname{section}{Section}{Section}

\renewcommand{\vec}[1]{\mathbf{#1}}
\newcommand{\mat}[1]{\mathbf{#1}}
\newcommand{\im}{\mathrm{j}}
\newcommand{\op}[1]{\mathcal{#1}}
\newcommand{\dd}{\mathrm{d}}

\newcommand{\citeNote}[1]{[\textcolor{orange}{X}]}

\tikzfading[name=fade l,left color=transparent!100,right color=transparent!0]
\tikzfading[name=fade r,right color=transparent!100,left color=transparent!0]
\tikzfading[name=fade d,bottom color=transparent!100,top color=transparent!0]
\tikzfading[name=fade u,top color=transparent!100,bottom color=transparent!0]

\definecolor{imw-blue}{rgb}{0.451,0.776,0.843}

\begin{document}

\onecolumn
\setcounter{page}{0}
\thispagestyle{empty}

\noindent\textbf{Author's Pre-Print}

\vspace{1em}

\noindent
This work has been submitted to the IEEE for possible publication. Copyright may be transferred without notice, after which this version may no longer be accessible.

\clearpage
\twocolumn

\setcounter{page}{1}

\title{Enhanced Wide-Angle Steering with Multi-Mode Multi-Port Aperture Antenna Arrays}

\author{
Tim Hahn\,\orcidlink{0000-0002-1421-3926}, \IEEEmembership{Graduate~Student~Member,~IEEE}, and Dirk Manteuffel\,\orcidlink{0000-0002-1490-2713},~\IEEEmembership{Member,~IEEE}

\thanks{This work was partly funded by the Federal Ministry of Research, Technology and Space of Germany within the project KOMSENS-6G under grant 16KISK127. \textit{(Corresponding author: Tim~Hahn.)}}
\thanks{The authors are with the Institute of Microwave and Wireless Systems, Leibniz University Hannover, 30167 Hannover, Germany (e-mail: \mbox{hahn@imw.uni-hannover.de}, \mbox{manteuffel@imw.uni-hannover.de}).}
}

\maketitle

\begin{abstract}
A novel concept for wide-angle scanning is proposed based on \aclp{M3PA}. The theory of \aclp{M3PA} based on aperture radiators is developed and applied towards the design of an antenna array consisting of multi-mode aperture radiators. An advanced beamforming algorithm is developed and implemented, making use of the higher degrees of freedom available to \aclp{M3PA}. The manufactured antenna array is measured and compared to the expected performance. Wide-angle steering up to $\pm77^\circ$ from broadside with respect to a scan loss of $\unit[3]{dB}$ is achieved in both the horizontal and vertical plane with no visible grating lobe.
\acresetall 
\end{abstract}


\section{Introduction}
\label{sec:introduction}

\IEEEPARstart{W}{ide-angle} beam steering has become a crucial capability in modern antenna systems, particularly with the rapid evolution of wireless technology for communication, sensing, radar and even next-generation technologies that aim to combine these features \cite{Li2023,Ghosh2025,Wild2023,Wild2021}.
A key technology enabler in this context are phased array antennas due to their flexibility, fast response, and ability to dynamically control high gain radiation patterns over a given field of view. 
However, conventional phased array systems based on patch antennas suffer from limited angular scanning performance within a typical scan range of $\pm 60^\circ$. Despite that, many emerging applications, such as millimeter-wave communications and satellite communication systems demand the ability to steer beams over larger angular ranges to achieve seamless spatial coverage and maintain reliable link performance \cite{Chaloun2022,Li2025}.

In recent years, wide-angle beam scanning has attracted significantly increased research interest, driven largely by the demands of beyond fifth-generation (B5G) and sixth-generation (6G) wireless systems \cite{Li2023}. These systems require high data rates, low latency, and efficient spectrum utilization, all of which benefit from wide spatial coverage and dynamic beam steering capabilities \cite{Li2023}. Additionally, satellite communications, automotive radar, and \ac{ISAC} systems can be enhanced by wide-angle scanning to improve coverage, target detection, and enable multi-user connectivity in highly dynamic environments \cite{Ghosh2025}.

The growing interest in wide-angle beam scanning is also accelerated by advances in enabling technologies, including metasurface-enhanced arrays \cite{Xi2021}, gradient index lens antennas \cite{Nie2023} and reconfigurable array elements \cite{Ahn2019}. These concepts allow antenna arrays to mitigate traditional limitations such as scan loss and scan blindness, which degrade performance at larger scan angles. Recent studies have demonstrated that array designs using pattern-reconfigurable antennas can achieve wide-angle scanning performance while maintaining stable beam characteristics \cite{Peng2021,Ding2017,Tang2026,Gao2019}. Higher order TM modes are typically used to create these reconfigurable elements, whereas \acp{CM} are a rather rarely used concept and occasionally used in post-processing for interpretation.

Our approach addresses these issues of conventional antenna concepts by using \acp{M3PA} to realize a pattern-reconfigurable antenna array with wide-angle scanning performance. The \ac{M3PA} elements are designed using \ac{CMA} to excite orthogonal currents, thereby ensuring decoupled ports. This concept systematically exploits the symmetry of the base radiator to generate different sets of orthogonal currents, while introducing a novel method to selectively excite these modes within an aperture radiator. Consequently, a key advantage of this \ac{CM}-based approach is that the antenna element ports are inherently decoupled without requiring additional decoupling effort.

\begin{figure*}
    \centering
    \includegraphics[height=7cm]{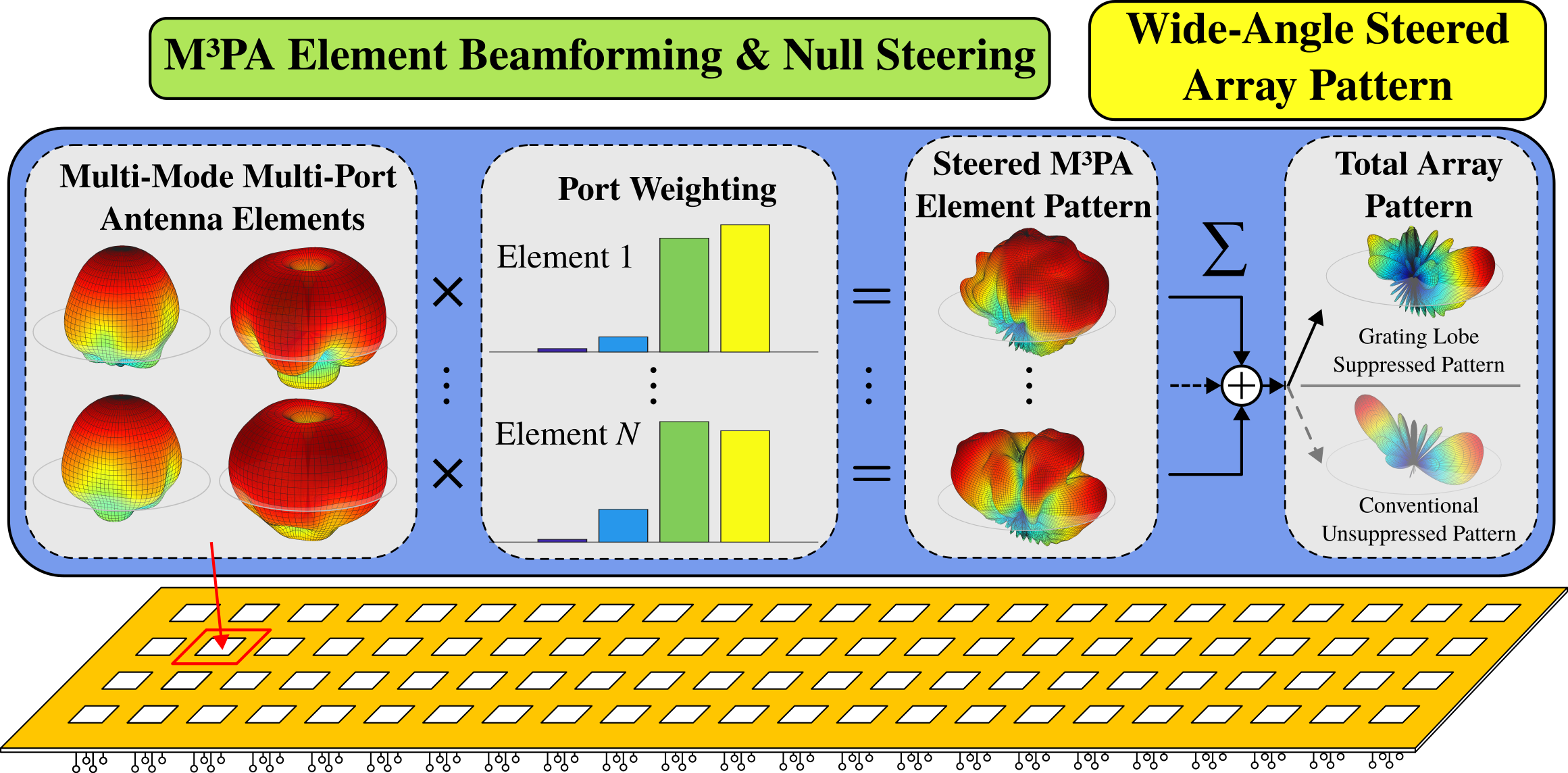}
    \caption{Schematic illustration of the proposed antenna array steering concept. The individual array pattern of the \ac{M3PA} elements are reconfigured in a first step to maximize the gain into the target steering direction, while simultaneously placing a null towards the direction of grating lobes. Then, all individual element patterns are summed up to obtain the array pattern of the \ac{M3PA} arrangement. The total array pattern reaches large steering angles angles without introducing grating lobes.}
    \label{fig:concept}
\end{figure*}

By enabling the excitation of multiple orthogonal currents and far-fields within the array elements, the individual array patterns can be reconfigured using weighting between the modes. For larger steering angles, modes with an off-broadside radiation characteristics are used, whereas the conventional broadside modes are used for broadside angles. This way, the reconfigured single element pattern can be used to enhance the total array pattern towards the desired direction and at the same time suppress any unwanted grating lobes. An overview of the basic procedure is illustrated in \cref{fig:concept}. 

The article is structured as follows. In \cref{sec:TCM}, an overview of the \acl{TCM} is given and the fundamental concepts required for the extension towards aperture problems are described and applied. In \cref{sec:beamforming}, a general overview on the concept of \aclp{M3PA} is shown and key characteristics of \acp{M3PA} are discussed. Furthermore, a beamforming algorithm is derived to make use of the higher degrees of freedom available to \acp{M3PA}. In this context, both analytical optimization methods as well as iterative optimizations are discussed. The theory and concepts are then applied within \cref{sec:design} to analyze and design the base element as well as the total \ac{M3PA} array operating in the \ac{mmwave} frequency range. Finally, measurements of the manufactured prototype are shown within \cref{sec:measurement} and compared to the expected performance.

\section{Theory of Characteristic Modes}
\label{sec:TCM}
\subsection{Theory of Characteristic Modes for Conducting Bodies}
The \ac{TCM} offers valuable insights into the physical radiation properties and radiation mechanisms of antennas. It is conventionally used to analyze the radiation properties of conducting bodies \cite{Harrington1971}. Herein, the \acp{CM} are defined based on the \ac{GEP} 
\begin{equation}
    \op{X} \left\{ \vec{J}_n \right\}= \lambda_n\op{R}\left\{ \vec{J}_n \right\}
    \label{eq:CM_GEP}
\end{equation}
formulated from the impedance operator $\op{Z}$ \cite{Harrington1971}, with $\op{R}$ and $\op{X}$ being the Hermitian and anti-Hermitian parts of the impedance operator respectively \cite{Harrington1971}. 
This concept of modes can be applied to decompose the total surface current density of an antenna into individual modes, revealing the radiation contributions of each mode to the total surface current density.
Thus, the total surface current density is the weighted sum of all modes of the structure \cite{Harrington1971}
\begin{equation}
    \vec{J} = \sum_{n} a_n \vec{J}_n = \frac{V_n^\text{i}}{1 + \im \lambda_n} \vec{J}_n.
\end{equation}
In this context, $V_n^\text{i}$ is the modal excitation coefficient. It describes how well the impressed electric field excites the $n$-th mode \cite{Harrington1971}.

More importantly, each \ac{CM} is associated with an eigenvalue $\lambda_n$, which characterizes the resonant behavior of the mode and indicates its proximity to resonance. Thus, the modal significance is a very important quantity and used to determine the resonance behavior of the individual modes over frequency. It is calculated from the eigenvalue $\lambda_n$ as \cite{CabedoFabres2007}
\begin{equation}
    \text{MS}_n = \frac{1}{|1+\im \lambda_n|}.
\end{equation}
Therefore, a low absolute eigenvalue corresponds to a high modal significance which indicates that the \ac{CM} contributes strongly to the total surface current density.

\subsection{Theory of Characteristic Modes for Apertures}
The concept of duality suggests, that the \ac{TCM} can be equally applied to aperture antennas. The theory has been studied in the past by researchers such as Harrington and Mautz \cite{Harrington1985,Harrington1976,Harrington1991,Kabalan1990,Kabalan2002,ElHajj1994,ElHajj1998}. More recently, design guidelines for slot antennas based on \acp{CM} have been formulated \cite{AntoninoDaviu2016,MohamedHicho2017,MohamedMohamedHicho2018} and the \ac{CMA} of aperture problems based on the duality principle has been proposed \cite{Liang2017,Liang2018}. However, the consequences and implications of the duality relations on the eigenvalue behavior of characteristic modes have not been mentioned explicitly. Therefore, a brief overview of the theory will be given here, as it is a prerequisite for the analysis and design within this work.

For the analysis of aperture radiators, the surface equivalence principle in combination with the method of images is commonly used to obtain the magnetic surface current distributions of dual aperture geometries \cite{Jin2010,Balanis2005}. From PMCHWT \cite{YlaeOijala2020}, the $\op{T}$ operator is known as
\begin{equation}
    \begin{split}
    \op{T}\left\{ \vec{F} \right\} =& \frac{\im}{k}\nabla\iint_{S'} G(\vec{r},\vec{r'}) \nabla' \cdot \vec{F}(\vec{r'})\,\dd S' \\&+ \im k \iint_{S'} G(\vec{r},\vec{r'}) \vec{F}(\vec{r'}) \, \dd S'
    \end{split}
    \label{eq:T_operator}
\end{equation}
where $\vec{F}$ describes either electric or magnetic source currents $\vec{J}_\text{s}$ and $\vec{M}_\text{s}$ respectively and the scalar free-space Green's function $G(\vec{r}, \vec{r'})$. Comparing the $\op{T}$ operator \cite{YlaeOijala2020} with the impedance and admittance operators from \cite{Harrington1971} and \cite{Harrington1976,Harrington1985}, the following relations can be identified given the duality of electric and magnetic currents
\begin{gather}
    \op{Z} = \left[\, \eta \, \op{T} \left\{ \vec{J}_\text{s} \right\} \,\right]_\text{tan} , \;
    \op{Y} = \left[\, 1/\eta \, \op{T}\left\{ \vec{M}_\text{s} \right\} \,\right]_\text{tan} \\
    \op{Y}\left\{\vec{M}_s \right\} \triangleq 1/\eta^2 \, \op{Z}\left\{\vec{J}_s \right\}
\end{gather}
with $\eta$ being the free-space impedance. 
As the \ac{GEP} from \eqref{eq:CM_GEP} is built upon the impedance operator, the similarity of impedance operator and admittance operator directly implies that the \ac{GEP} for apertures
\begin{equation}
    \op{B} \left\{ \vec{M}_n \right\}= \lambda_n \op{G} \left\{ \vec{M}_n \right\}
\end{equation}
is a scaled variant of the dual \ac{GEP} for conducting bodies. 
Consequently, the eigenvalues and electric and magnetic eigencurrents are equal for a conducting body and its dual aperture counterpart which can be formulated as
\begin{equation}
    \lambda_{n,\text{PEC}} \triangleq \lambda_{n,\text{Aperture}} \; , \text{with} \; 
    \vec{J}_{n,\text{PEC}} \triangleq \vec{M}_{n,\text{Aperture}}.
\end{equation}
Leveraging this duality relationship thus provides an elegant and systematic foundation for the design of complex aperture radiators using \ac{TCM}.

\section{Advanced Beamforming Using Multi-Mode Multi-Port Antenna Arrays}
\label{sec:beamforming}
\subsection{Multi-Mode Multi-Port Antennas}
\Acfp{M3PA} are a special kind of antenna type which aim to selectively excite different sets of characteristic modes on a shared antenna structure using multiple ports. We have been working on these advanced antenna types for many years and built a comprehensive theory for the analysis and design of such antennas \cite{Manteuffel2016,Peitzmeier2019,Peitzmeier2022}.
Key characteristics of these \acp{M3PA} include low port coupling as well as spatially diverse radiation patterns with low \acp{ECC} which makes them especially suitable for \ac{MIMO} applications. The conceptual structure of an \ac{M3PA} includes the following components. The antenna element on which the \acp{CM} are excited, multiple feeds or exciters distributed over the antenna element, as well as a feed network or integrated circuit to excite the feeds with their required amplitude and phase. 

Regarding the analysis of \acp{M3PA}, it has been discovered that the \acp{CM} of a structure are closely linked to its symmetry properties. In this context, group theory has been identified to be a valuable tool to describe the orthogonality properties and eigenvalue behavior of \acp{CM} on symmetric structures \cite{Peitzmeier2022,Peitzmeier2019,Peitzmeier2019a,Peitzmeier2019b,Masek2020,Masek2019,Schab2017}. One of the main outcomes of group theory applied towards the \ac{TCM} is that the maximum number of realizable orthogonal ports on a \ac{M3PA} is limited by the symmetry order of the antenna structure. Consequently, higher numbers of orthogonal ports can be realized on an antenna by increasing its symmetry order \cite{Peitzmeier2019}. 

As single element \acp{M3PA} can be flexibly reconfigured, there is significant potential for the application of \acp{M3PA} in antenna arrays \cite{Peitzmeier2018,Hahn2024a}. Using reconfigurable antenna elements in complete arrays enables higher degrees of freedom for the configuration of these arrays compared to conventional single port antennas. 

\subsection{Beamforming Optimization Algorithm}
\label{sec:bf_algorithm}
To enable a wide-angle scanning range, our approach is to use \acf{M3PA} radiators with reconfigurable element patterns that can be dynamically optimized with respect to the desired scan angle of the array. While conventional arrays can be easily described using pattern multiplication to synthesize total array patterns \cite{Balanis2005,Mailloux2005}, the situation changes for pattern-reconfigurable antennas as these have multiple radiation patterns available to individually choose from per element of the array.

In order to utilize these higher degrees of freedom available to them, a suitable array beamforming algorithm needs to be implemented. Several algorithmic strategies exist in literature for beamforming with pattern‑reconfigurable elements. A common approach combines discrete mode/state selection with classical amplitude/phase weighting, where joint mode-selection and excitation optimization is performed using heuristic optimizations \cite{Ding2017}. Other authors implement closed-form solutions to synthesize the weights of the reconfigured array elements \cite{Wang2024,Deshpande2026}. We choose a manifold optimization based approach similar to \cite{Salmi2024a} due to its computational efficiency, flexibility and deterministic robustness compared to the aforementioned.
Therefore, an array beamforming algorithm is implemented based on a conjugate gradient optimizer. The algorithm is implemented using the \textit{Manopt toolbox} \cite{manopt}.

To start the optimization process, a desired far-field pattern is initialized. For this purpose, a pencil beam with adjustable beam width is created within the target angular sector
\begin{equation}
    \mat{E}_\text{d}(\theta,\varphi) = 
    \begin{cases}
        E_0\, \mat{e}_\text{pol} , & \theta \in \left[ \theta_\text{d,min}, \theta_\text{d,max} \right], \varphi \in \left[ \varphi_\text{d,min}, \varphi_\text{d,max} \right]\\
        0, & \text{else.}
    \end{cases}
\end{equation}
The amplitude $E_0$ of the desired pattern is chosen such that the radiated power within the target sector is normalized to
\begin{equation}
    P_\text{rad,d} = \oiint\limits_{S} \frac{1}{2 \eta} |E_0|^2 \mat{e}_\text{pol} \, \dd \mathbf{S} = \unit[1]{W}.
\end{equation}
Further, the polarization vector $\mat{e}_\text{pol}$ is chosen to be
\begin{equation}
    \mat{e}_\text{pol} = 
    \begin{pmatrix}
        \cos(\psi)\\
        \sin(\psi)
    \end{pmatrix}
    ,\qquad \psi \in \left[0,2\pi \right]
\end{equation}
with $\psi = 0$ resulting in $\theta$-polarization and $\psi=\pi$ resulting in $\varphi$-polarization.
The target polarization needs to be defined before the optimization procedure begins.
The optimization algorithm shall maximize the \acf{ECC} between the desired pattern and the weighted embedded element pattern superposition
\begin{equation}
    \max_{\mathbf{a} \in \mathbb{R}^{N}} \mathrm{ECC}\!\left\{\mat{a} \cdot \mat{E}_\text{p},\, \mat{E}_\text{d} \right\}
\end{equation}
with the $\mat{a}=(a_1,\dots,a_N)$ being the coefficients of the $N$ array ports and the \ac{ECC} operator being defined as \cite{Safin2013}
\begin{equation}
    \mathrm{ECC}\!\left\{ \mat{E}_\alpha, \mat{E}_\beta \right\} = \frac{1}{2 \eta}
    \oiint\limits_{S} \frac{ \mat{E}_\alpha \cdot  \mat{E}_\beta^\ast}{\sqrt{P_{\text{rad},\alpha}} \, \sqrt{P_{\text{rad},\beta}}} \dd S.
    \label{eq:ECC}
\end{equation}
For the optimization with non-uniform coefficient amplitudes, a complex sphere manifold is used with the cost function being defined as
\begin{equation}
    f_\text{cost}(\mat{a}) = - \mathrm{ECC}\!\left\{\mat{a} \cdot \mat{E}_\text{p},\, \mat{E}_\text{d} \right\}
\end{equation}
and its derivative with respect to the port coefficients $\mat{a}$
\begin{equation}
    \frac{\partial f_\text{cost}(\mat{a})}{\partial \mat{a}} = -\mathrm{ECC}\!\left\{\mat{E}_\text{p},\, \mat{E}_\text{d} \right\}.
\end{equation}
Thus, a maximum correlation between the target pattern and the optimized array pattern yields minimum costs for the optimizer.
Consequently, the optimizer minimizes the costs for all variations of the port coefficients
\begin{equation}
    \min_{{\mathbf{a} \in \mathbb{R}^{N}}} \! f_\text{cost}(\mat{a}).
\end{equation}
The procedure requires all far-field patterns of the array ports to be present (either from simulation or measurement) for this method to work. Further, if embedded element patterns are used, coupling between neighboring elements is incorporated in the calculations of the total array pattern.

In case uniform amplitude distributions are of interest, instead a complex circle manifold can be used and the cost function and its derivative are scaled by a factor $1/\sqrt{N}$ to account for port power normalization.
In practice, the port weights only need to be computed once for each target beam direction and can be stored in a lookup table for low-latency reuse.
The optimization algorithm will be used and applied towards the array design described in the following chapter.

\section{Multi-Mode Multi-Port Antenna Array Design}
\label{sec:design}
\subsection{Antenna Element Analysis and Design}
\label{sec:element_design}
To design a \ac{M3PA} antenna array, the element radiator needs to be analyzed and designed systematically to identify the modes of the structure and to determine potential feed structures for these modes. The element radiator shall enable the excitation\footnote{To be precise, eigencurrents cannot be excited directly in general. Instead, currents exhibiting characteristics similar to those of the corresponding eigencurrents are excited. This interpretation of excitation is adopted throughout the remainder of this article.} of different orthogonal sets of \acp{CM} by different antenna ports.

For the specific aperture-based \ac{M3PA} radiator designed within this work, the millimeter wave band n257 ranging from $\unit[26.5]{GHz} \leq f \leq \unit[29.5]{GHz}$ with a center frequency of $f_\text{c}=\unit[28]{GHz}$ is chosen. 

As a base structure of the \ac{M3PA} element, a square aperture is chosen, as it is known to possess six orthogonal current groups \cite{Peitzmeier2019} from which the desired modes can be chosen. A square aperture is advantageous for \ac{M3PA} array design, as its geometry aligns naturally with square lattices and minimizes unused space between elements. Moreover, it provides four \acp{CM} that can be excited using a single feed network \cite{Hahn2024}, requiring only four feeds and thus keeping the implementation effort reasonable.

As known from array theory, the footprint of array elements should stay low to have the antenna elements positioned closely together to avoid the occurrence of grating lobes in the visible region when steering. At the same time, the \ac{M3PA} elements need to be large enough to have higher order modes significant. In general, these two properties are in conflict with each other. 

To increase the modal significance of higher order modes while maintaining a low footprint of the antenna element, the aperture is filled with a dielectric substrate with $\varepsilon_\text{r,Sub.}>1$ to shift the mode resonances towards lower frequencies. However, this insertion of the dielectric creates an impedance mismatch between the aperture and free-space due to the higher effective permittivity of the filled aperture. To mitigate this problem, a tapered superstrate with $\varepsilon_\text{r,Sup.}\approx\varepsilon_\text{r,Sub.}$ is placed on top of the filled aperture. The superstrate acts as an impedance taper, gradually transitioning the mode impedance from the substrate-filled aperture to free-space, thereby minimizing the impedance mismatch. The influence of the superstrate on the steering performance is investigated in the \hyperref[appendix]{Appendix}.

\begin{figure}
    \centering
    \input{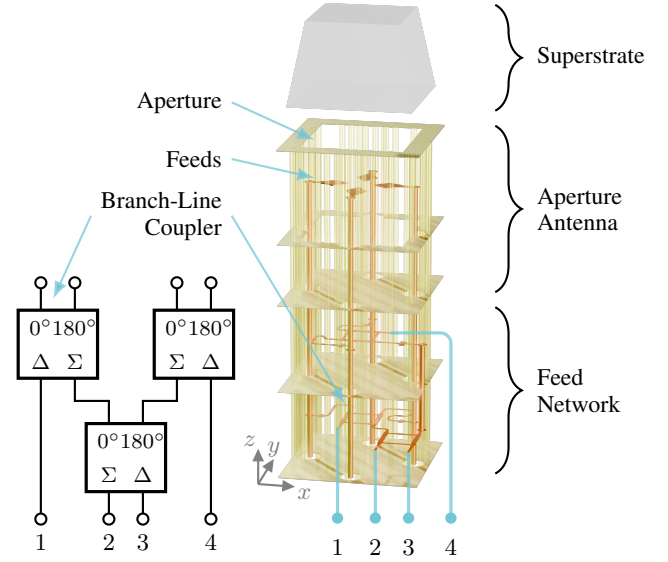}    
    \caption{Exploded view of the antenna integrated into a multi-layer \acf{PCB}. The radiating aperture is covered by a superstrate to improve the impedance transition to free-space. The feeds are embedded into the \ac{PCB} and fed by a feed network. The antenna aperture and all feed components are surrounded by shielding vias.}
    \label{fig:antenna_concept}
\end{figure}

The conceptual drawing of the antenna element is shown in \cref{fig:antenna_concept}.
The antenna is realized in \acf{PCB} technology with the radiating antenna aperture positioned on the top layer and excitation feeds embedded into the \ac{PCB}. The feeds are excited by a feed network with their corresponding amplitude and phase to selectively excite modes of different orthogonal current groups. The feed network consists of branch-line couplers and is distributed across multiple layers to limit its dimensions to approximately the aperture size. The superstrate with a dielectric constant of $\varepsilon_{r,\text{Sup.}}=4.5$ is located on top of the substrate-filled aperture. The superstrate has a pyramid shape to introduce an air gap between adjacent elements to prevent the propagation of substrate waves inside the superstrate. It has a bottom base area of $A_\text{Sup.,bot} = \unit[6]{mm} \times \unit[6]{mm}$ and top area of $A_\text{Sup.,top} = \unit[4.5]{mm} \times \unit[4.5]{mm}$ with a height of $h_\text{Sup.}=\unit[4]{mm}$ and a linear taper from bottom to top.

\begin{figure}
    \centering
    \includegraphics[width=\linewidth]{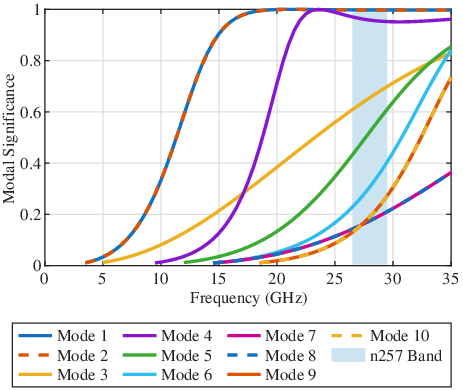}
    \caption{Modal significances of the square aperture with dimensions \mbox{$A \approx 0.467 \lambda_0 \times 0.467 \lambda_0 \approx 0.74 \lambda_\text{eff} \times 0.74 \lambda_\text{eff} = \unit[5]{mm} \times \unit[5]{mm}$}.}
    \label{fig:modal_significance}
\end{figure}

To reveal the modes inherent to the square aperture, a \ac{CMA} of the square aperture is carried out. For the analysis of the magnetic \acp{CM} within the aperture and their eigenvalue behavior over frequency, the aperture dimensions are calculated within an effective medium, as the aperture is embedded into the dielectric material. Therefore, the average dielectric permittivity of substrate and superstrate is determined and used to calculate the effective permittivity as a first order approximation as \cite{Sorrentino2010}
\begin{equation}
    \varepsilon_{\text{r,eff}} = \frac{(\varepsilon_{\mathrm{r,Substrate}} + \varepsilon_{\mathrm{r,Superstrate}})/2+1}{2}.
\end{equation}
The relative permittivity of the substrate used for the determination of the effective permittivity is $\varepsilon_{r,\text{Sub.}}=3.62$.
The modal significances calculated in this way for the equivalent magnetic aperture currents embedded in the effective medium are shown in \cref{fig:modal_significance}.
\begin{figure}
    \includegraphics{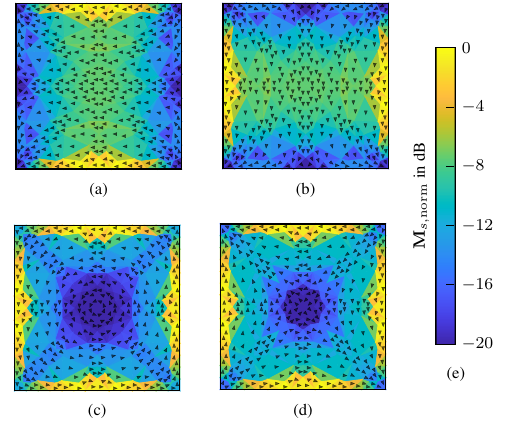} 
    \caption{Characteristic magnetic surface current densities $\vec{M}_{s,n}$ of the first four eigenmodes of the square aperture. (a)~\mbox{Dipole mode 1.} (b)~\mbox{Dipole mode 2.} (c)~Loop mode. (d)~Quadrupole mode. (e)~Legend.
    }
    \label{fig:cms_square}
\end{figure}
As can be seen from \cref{fig:modal_significance}, modes $1-4$ have the highest modal significance inside the target frequency range and are thus considered for excitation. Modes $1$ and $2$ form a degenerate mode pair and thus share the same eigenvalue trace. Due to the symmetry of the square aperture, modes $1-6$ are all orthogonal to each other. In this context, orthogonality is defined as the current correlation coefficient between the currents yielding zero
\begin{equation}
    \rho_{uv} = \frac{\oiint_S \vec{M}_u \cdot \vec{M}_v^\ast \,\dd S}{\sqrt{\oiint_S |\vec{M}_u|^2\,\dd S}\sqrt{\oiint_S |\vec{M}_v|^2\,\dd S}} = \delta_{uv}
\end{equation}
with $\delta_{uv}$ being the Kronecker delta.

The first four characteristic magnetic current modes of the aperture in the effective medium are shown in \cref{fig:cms_square}. The modes $1$ and $2$ correspond to the dipole modes, which excite horizontally- and vertically-polarized far-field patterns, respectively. Here, horizontal polarization is defined along the $x$-direction and vertical polarization along the $y$-direction, with respect to the coordinate system shown in \cref{fig:antenna_concept}. Modes $3$ and $4$ are higher order modes, with mode $3$ being referred to as loop mode and mode $4$ being referred to as quadrupole mode. Both of the higher order modes excite far-fields with mixed polarization, a broadside null, and off-broadside radiation characteristics. Thus, the higher order modes are advantageous when steering towards large angles, whereas the conventional dipole modes are better suited to steer towards broadside.

\begin{figure}
    \centering
    \input{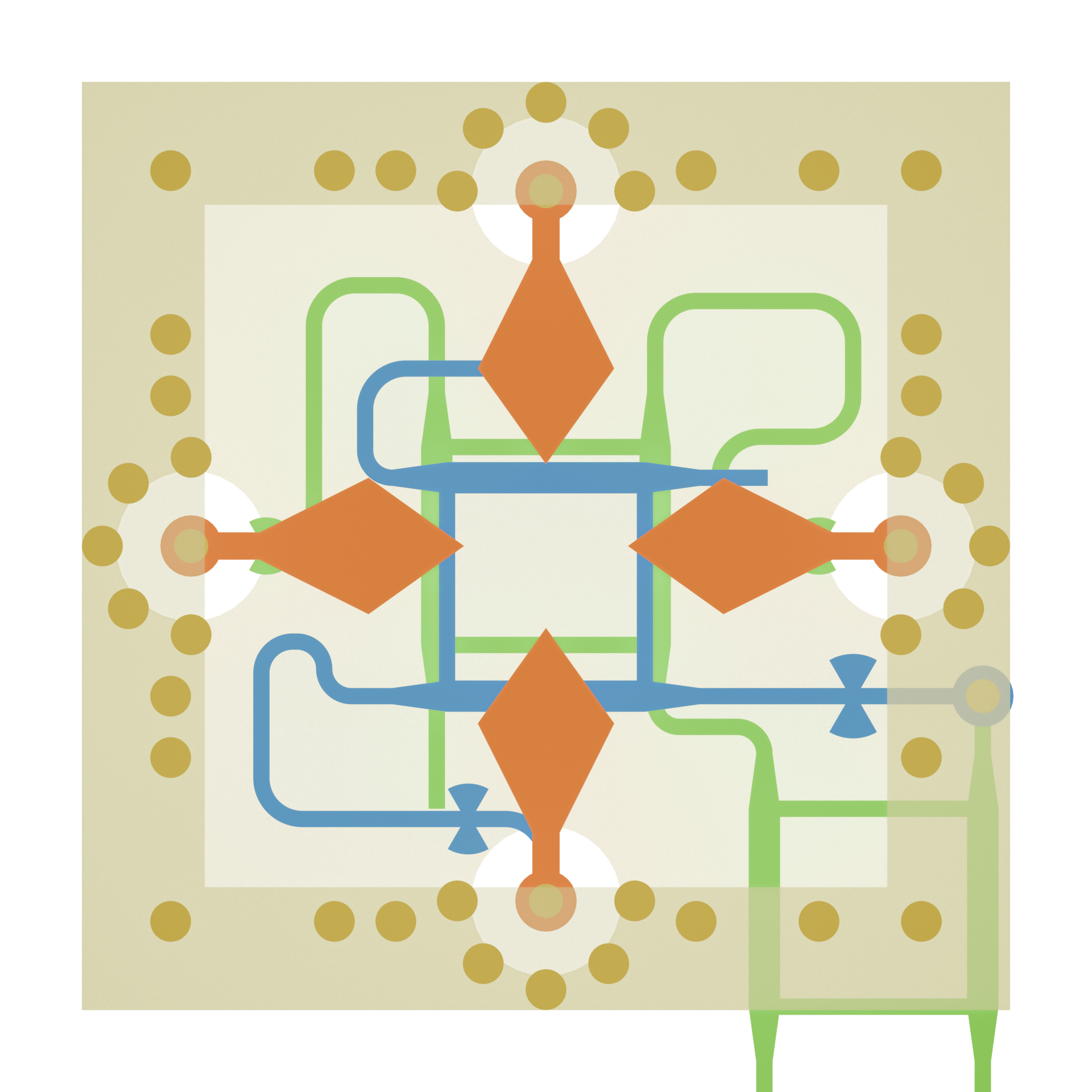}
    \caption{Feed structure of the element radiator. The feeds are embedded into the \ac{PCB} with the feed networks located on lower layers. The feed network is connected to the input of the feeds using via transitions. The total aperture is surrounded by via walls to establish the boundary conditions for the aperture mode. The feeds are separated from the networks by a ground plane. They are located at a height of $h=\unit[1.29]{mm}\approx \lambda_\mathrm{eff}/4$ above the ground plane. The dimensions of the feeds are as follows: $a=\unit[5]{mm}$, $b=\unit[0.8]{mm}$, \mbox{$c=\unit[0.7]{mm}$}, $d=\unit[0.2]{mm}$, $w=\unit[1]{mm}$, $l=\unit[2]{mm}$, $w_\mathrm{Line} = \unit[0.12]{mm}$.}
    \label{fig:feeds}
\end{figure}

The feed structure used to excite the modes in \cref{fig:cms_square} is shown in \cref{fig:feeds}. The excitation elements are essentially monopoles that are arranged in a square. The feed elements are located below to uppermost radiating square aperture layer. The feeders have a distance of a quarter wavelength to the ground plane beneath them in order to maximize the constructive interference with the reflected wave from the ground plane. The inputs of the feed elements are connected to the outputs of the feed network which is located on the lower layers of the \ac{PCB}, as shown in \cref{fig:concept}. 
The feed network provides the antenna feeds with the required amplitude and phase excitations.
It is realized using cascaded branch-line couplers in stripline technology with $90^\circ$ delay lines attached at the outputs to realize the total required phase shift of $\Delta\varphi = 180^\circ$. The delay line is necessary, as the branch-line coupler on its own would only be able to realize a $90^\circ$ phase shift. Although a rat-race coupler satisfies the feed network requirements, its stripline implementation would require trace widths below $\unit[100]{\mu m}$, posing manufacturing challenges. Therefore, a branch-line coupler was selected due to its wider trace widths and improved manufacturability in \ac{PCB} technology.

To enable the transition from the feed network layers to the feeds, the feeds are surrounded by vias to support the propagation of a quasi-coaxial mode. At these high frequencies, matching stubs are incorporated to optimize the transition impedance and bandwidth. In order to realize the boundary conditions required for the aperture, the aperture is surrounded by through-hole vias. These surrounding via walls enable the excitation of waveguide modes within the cavity and furthermore act as shielding vias within neighboring antenna elements in the array. This way, the coupling between the array elements can be reduced by integrating the shielding technique directly into the antenna element.

\subsection{Array Synthesis, Design and Implementation}
\label{sec:array_design}
To achieve symmetrical wide-angle steering for both the horizontal and vertical plane, the array lattice should also be chosen symmetrically w.r.t. both planes. As the array element radiator has been designed systematically to excite \acp{CM} of different orthogonal current groups, the obvious choice for the array lattice is also a square geometry, where the elements are equally spaced in $x$- and $y$-direction w.r.t. the coordinate system shown in \cref{fig:feeds}. An advantage of this array topology is that the antenna aperture per element radiator can be maximized while minimizing unused space between the elements, as the square shape naturally fits the array lattice. This consequently leads to higher gains available to the element.

Due to the project specifications requiring a minimum realized gain of $\unit[16]{dBi}$ in broadside direction, a $4\times4$ element array with an inter-element spacing of $d=\unit[6.5]{mm}\approx0.6\,\lambda_0$ has been chosen. The maximum realized gain of the total array structure can be estimated from the broadside single element pattern using
\begin{equation}
    G_\mathrm{Array,max} \approx 10\log_{10} (N\cdot M)\,\mathrm{dB} + G_\mathrm{Elem.}.
\end{equation}
This yields a maximum array gain of $G_\mathrm{Array,max}\approx\unit[19.7]{dBi}$ for the $4\times4$ array. Thus, the array provides a sufficient margin with respect to the specifications. 

The total number of antenna array ports for the $4\times4$ array with $4$ ports per radiator is $N_\mathrm{Ports} = 4\times4\times4=64$, providing a high degree of freedom and flexibility for reconfiguring the overall antenna array. For an industrialized antenna array, minimizing the total number of ports is generally desirable. However, the objective of this work is to investigate the additional degrees of freedom enabled by \ac{M3PA} arrays. Therefore, assigning a dedicated port to each element maximizes reconfigurability and facilitates exploration of the performance limits of the proposed array architecture.

\begin{figure}
    \centering
    \input{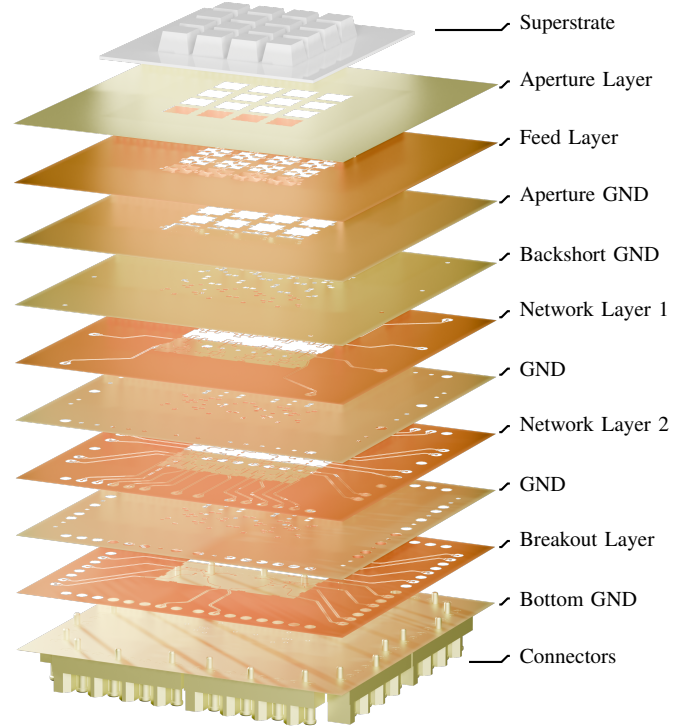}
    \caption{Exploded view of the \ac{M3PA} array integrated into the multi-layer \ac{PCB}. The \ac{PCB} has a total number of ten layers. Connectors for distributing the input signals into the \ac{PCB} are soldered to the bottom side of the \ac{PCB}. Vias are hidden in this illustration to increase the visibility of layer structures.
    }
    \label{fig:array_exploded}
\end{figure}

Using the synthesized parameters, the total antenna array is designed based on the single element radiator described in \cref{sec:element_design}. An exploded view of the \ac{M3PA} array stack-up in multi-layer \ac{PCB} technology is shown in \cref{fig:array_exploded}. The \ac{PCB} consists of a total number of ten layers with the transmission lines integrated in stripline technology. The radiating antenna aperture is located on the highest layer of the \ac{PCB} with the excitation feeds of the antenna element located on the layers beneath it. A breakout layer is positioned beneath the feed networks to route and distribute the input signals to the individual antenna elements. The input signals are fed to the \ac{PCB} using $8\times1$ connectors from the Rosenberger Mini-Coax Multi-Port series, which are located on the bottom side of the \ac{PCB}.

In total, the \ac{PCB} has a footprint of $A_\mathrm{PCB} = \unit[60]{mm}\times\unit[60]{mm}$ with a thickness of $t_\mathrm{PCB} = \unit[2.616]{mm}$. As a core material for the \ac{PCB} cores, Panasonic Megtron 6 R-5775(K) is used with a dielectric constant of $\varepsilon_r = 3.62$ and dissipation factor of $\tan(\delta)=0.0052$ at $\unit[24]{GHz}$. The superstrate is printed in a \ac{FDM} 3D-printer using the Zetamix-$\varepsilon$-$4.5$ filament with a dielectric constant of $\varepsilon_r = 4.5$ and a dissipation factor of $\tan(\delta)=0.0015$. 

\section{Prototype, Measurement and Validation}
\label{sec:measurement}
\subsection{Measurement and Validation of the Single Element Antenna}
\begin{figure}
    \centering
    \subfloat[]{
    \includegraphics[width=0.25\linewidth]{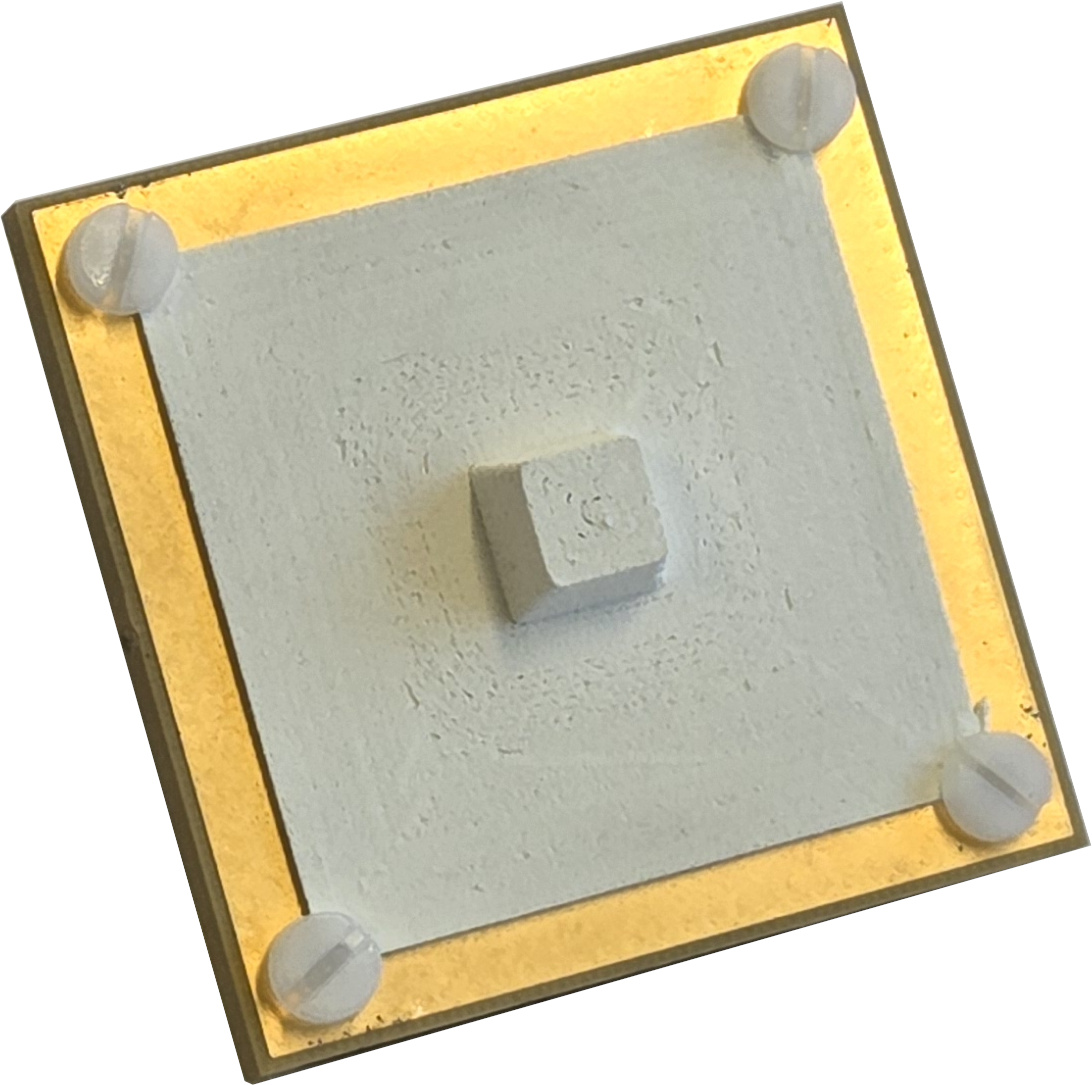}
    }
    \hspace{1.5cm}    
    \subfloat[]{
    \includegraphics[width=0.25\linewidth]{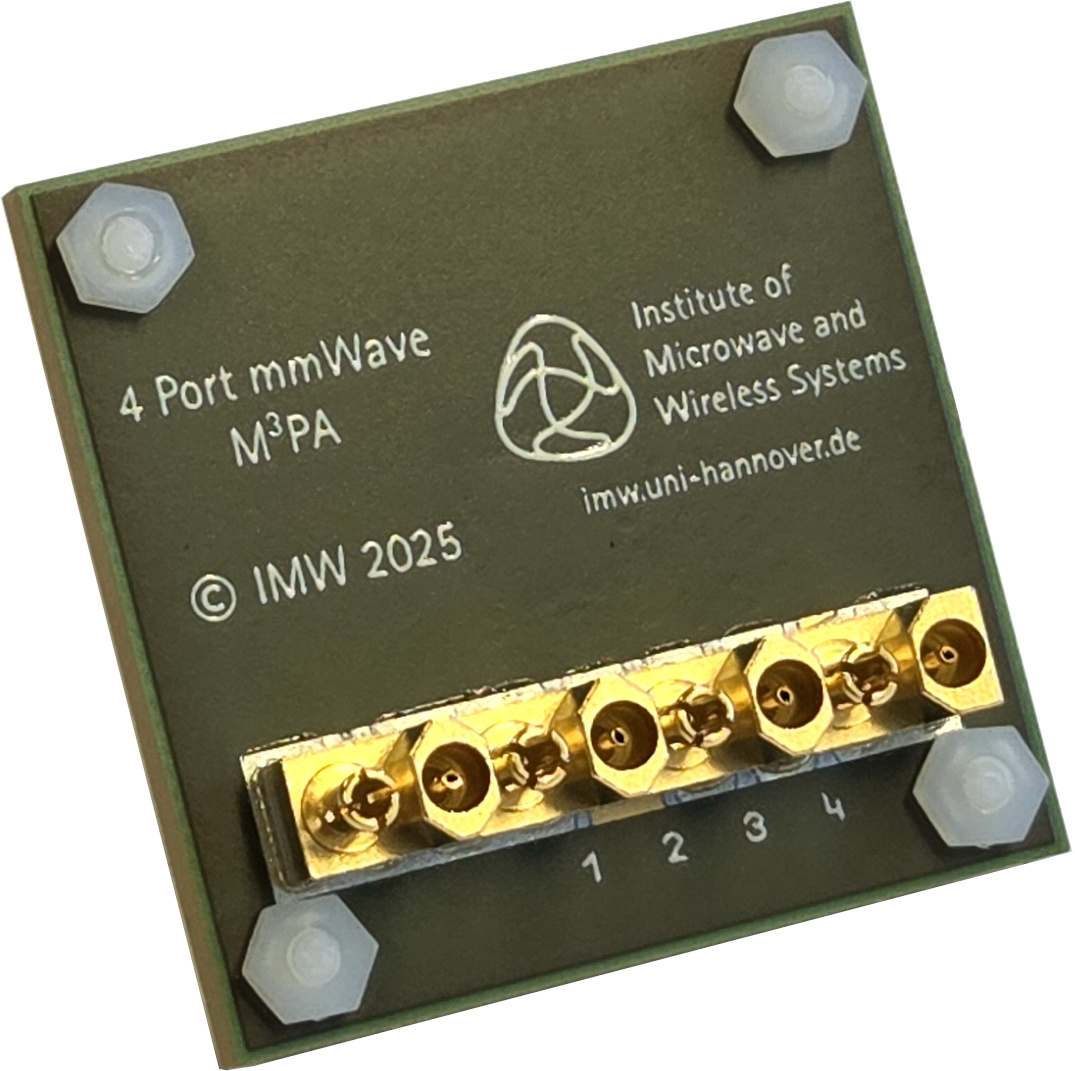}
    }
    \\
    \subfloat[]{
    \includegraphics[width=0.45\linewidth]{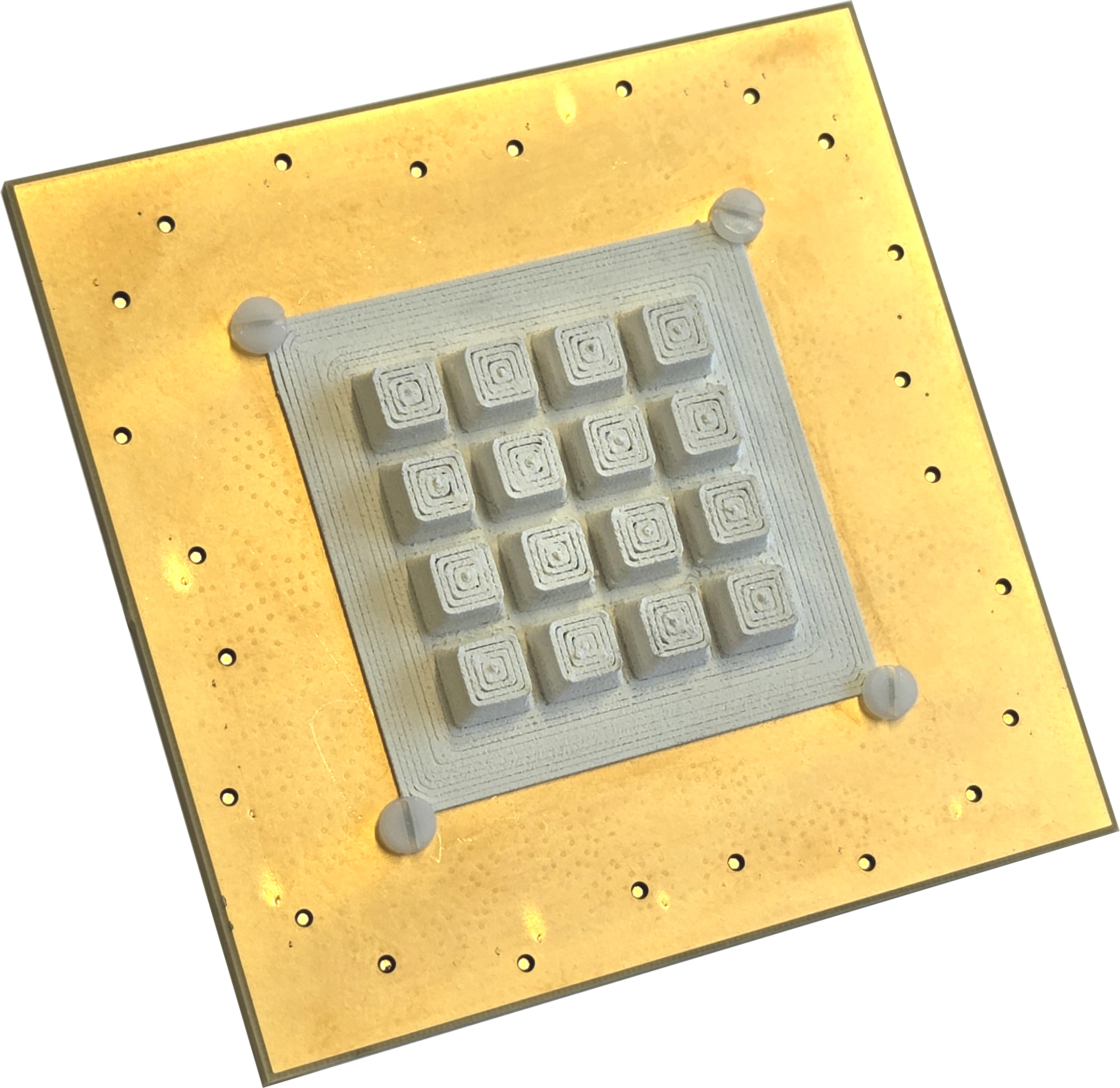}
    }
    \subfloat[]{
    \includegraphics[width=0.45\linewidth]{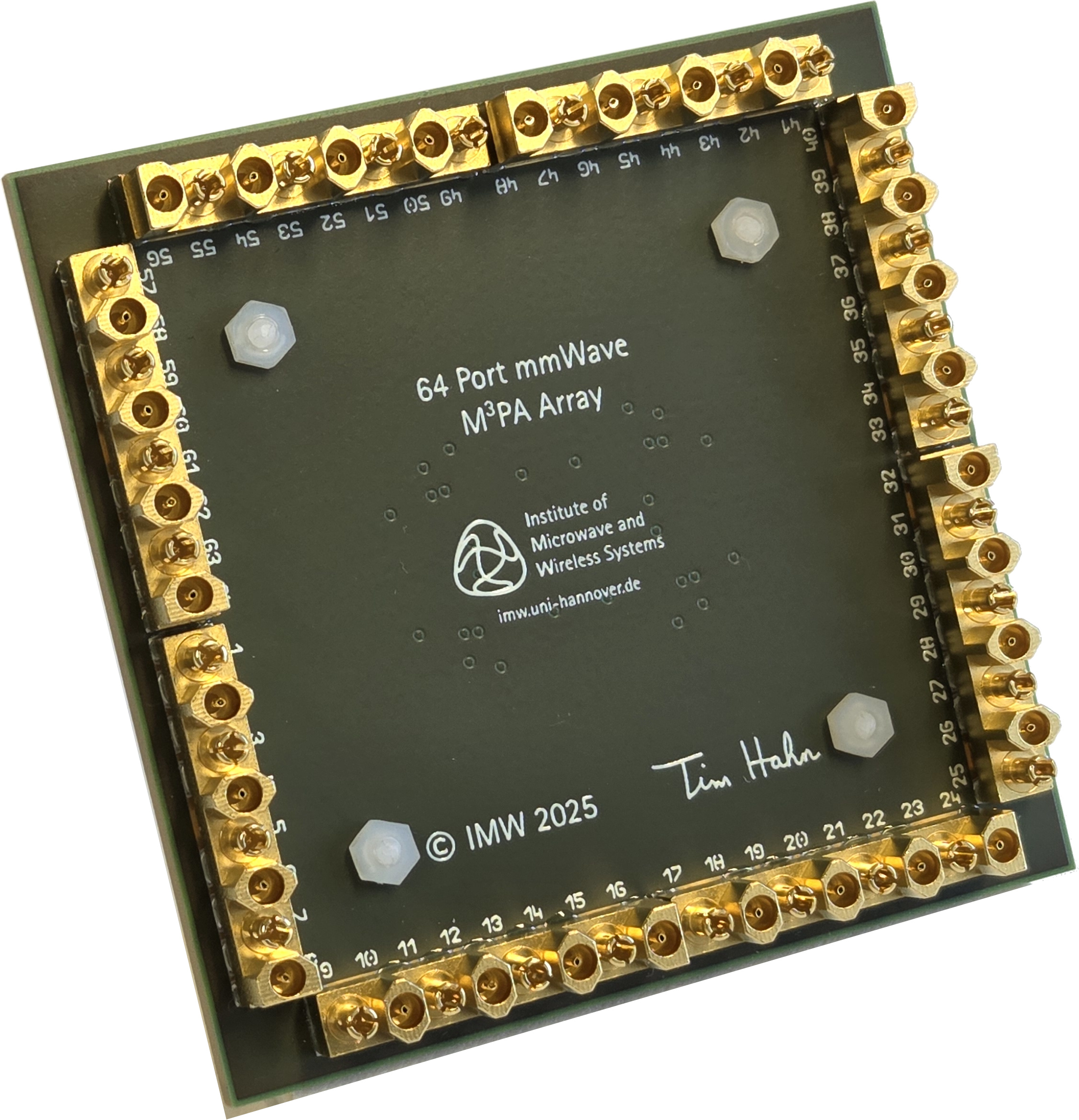}
    }
    \caption{Front and back view of the manufactured single antenna prototype and $4\times4$ antenna array prototype. Both prototypes are manufactured in \ac{PCB} technology and incorporate signal distribution lines as well as a feed network. Connectors for the individual antenna ports are placed on the bottom side of the \acp{PCB}.}
    \label{fig:manufactured_prototypes}
\end{figure}

\begin{figure}
    \centering
    \subfloat[\label{fig:spar:p1}]{
    \includegraphics[]{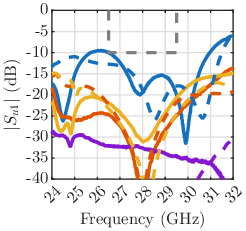}
    }
    \subfloat[\label{fig:spar:p2}]{
    \includegraphics[]{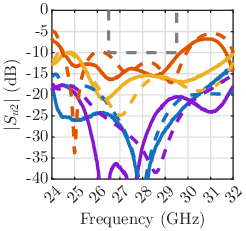}
    }\\
    \subfloat[\label{fig:spar:p3}]{
    \includegraphics[]{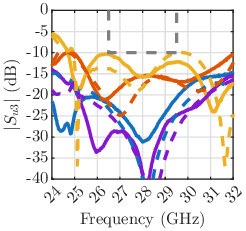}
    }
    \subfloat[\label{fig:spar:p4}]{
    \includegraphics[]{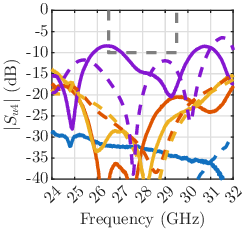}
    }\\
    \subfloat[\label{fig:spar:legend}]{



\begin{tikzpicture}[every node/.style={font=\footnotesize}]

        \newcommand{\figW}{6.8}
        \newcommand{\figH}{0.9}
        \newcommand{\xA}{0.3}
        \newcommand{\yA}{0.65}
        \newcommand{\yB}{0.25}
        \newcommand{\wBox}{0.6}
        \newcommand{\hBox}{0.3}
        \newcommand{\dist}{1.8}
        \newcommand{\distB}{1.2}
        \newcommand{\lineW}{1}

        \draw[thick] (0,0) rectangle (\figW,\figH);

        \fill[fill={rgb,255:red,18;green,112;blue,191}] (\xA,\yA-0.5*\hBox) rectangle (\xA+\wBox,\yA+0.5*\hBox);
        \node[right] at (\xA+\wBox,\yA) {$u = 1$};    

        \fill[fill={rgb,255:red,222;green,84;blue,0}] (\xA+\dist,\yA-0.5*\hBox) rectangle (\xA+\wBox+\dist,\yA+0.5*\hBox);
        \node[right] at (\xA+\wBox+\dist,\yA) {$u = 2$}; 

        \fill[fill={rgb,255:red,237;green,176;blue,33}] (\xA,\yB-0.5*\hBox) rectangle (\xA+\wBox,\yB+0.5*\hBox);
        \node[right] at (\xA+\wBox,\yB) {$u = 3$};    

        \fill[fill={rgb,255:red,132;green,23;blue,209}] (\xA+\dist,\yB-0.5*\hBox) rectangle (\xA+\wBox+\dist,\yB+0.5*\hBox);
        \node[right] at (\xA+\wBox+\dist,\yB) {$u = 4$};   

        \draw[gray, line width=1.5pt] (\xA+\wBox+\dist+\distB,\yA) -- (\xA+\wBox+\dist+\distB+\lineW,\yA);
        \node[right] at (\xA+\wBox+\dist+\distB+\lineW,\yA) {Measurement};

        \draw[gray, line width=1.5pt, dash pattern=on 4pt off 4pt] (\xA+\wBox+\dist+\distB,\yB) -- (\xA+\wBox+\dist+\distB+\lineW,\yB);
        \node[right] at (\xA+\wBox+\dist+\distB+\lineW,\yB) {Simulation};

    \end{tikzpicture}

    }
    \caption{Measured and simulated S-parameters $|S_{uv}|$ of the fabricated single element \ac{M3PA}. (a)~Port 1 ($v=1$). (b)~Port 2 ($v=2$). (c)~Port 3 ($v=3$). (d)~Port 4 ($v=4$). (e)~Legend.}
    \label{fig:spar_elem}
\end{figure}
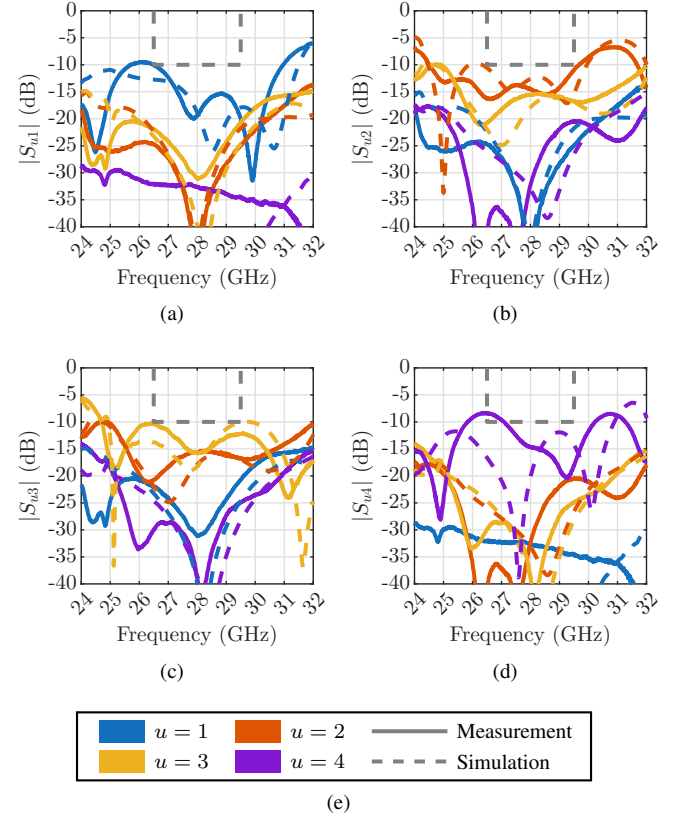

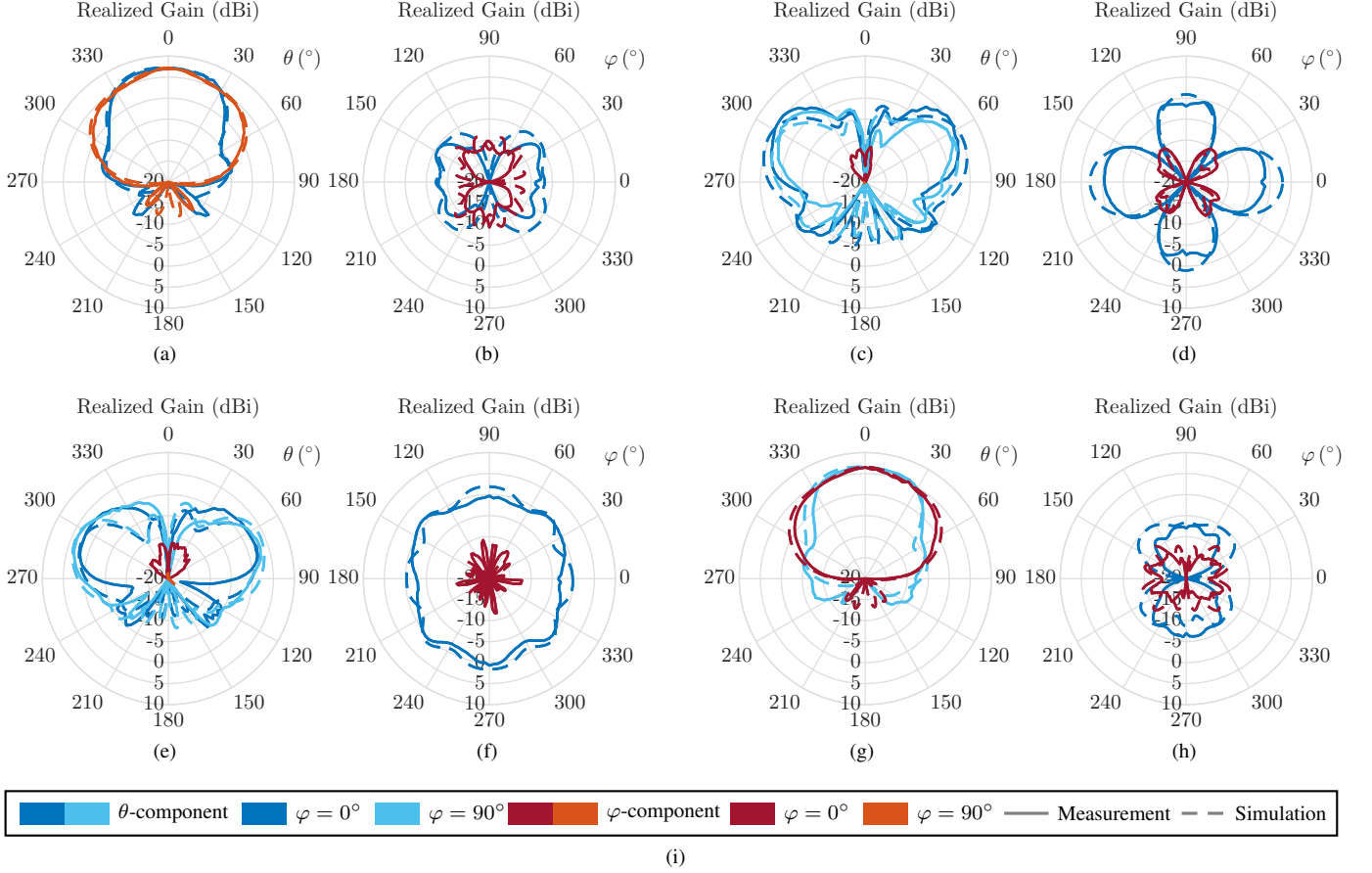
\begin{figure*}
    \centering
    \subfloat[\label{fig:ff_elem:a}]{\includegraphics[]{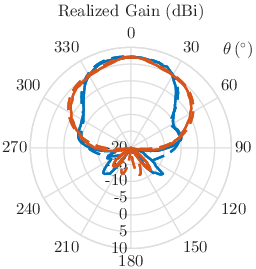}}
    \subfloat[\label{fig:ff_elem:b}]{\includegraphics[]{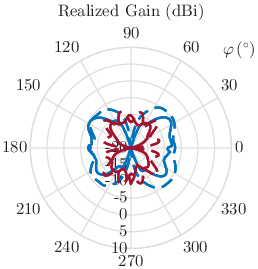}}
    \hfill
    \subfloat[\label{fig:ff_elem:c}]{\includegraphics[]{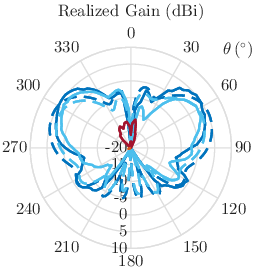}}
    \subfloat[\label{fig:ff_elem:d}]{\includegraphics[]{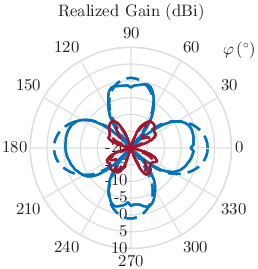}}\\
    \subfloat[\label{fig:ff_elem:e}]{\includegraphics[]{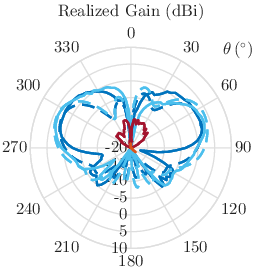}}
    \subfloat[\label{fig:ff_elem:f}]{\includegraphics[]{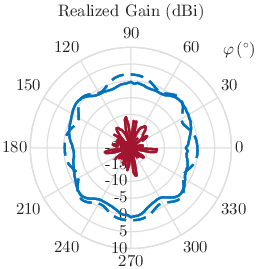}}
    \hfill
    \subfloat[\label{fig:ff_elem:g}]{\includegraphics[]{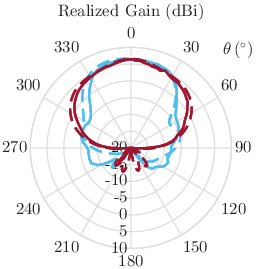}}
    \subfloat[\label{fig:ff_elem:h}]{\includegraphics[]{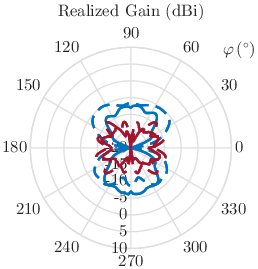}}\\
    \subfloat[\label{fig:ff_elem:leg}]{



\begin{tikzpicture}

\def\dist{6.6}
\def\xA{0.2}
\def\boxW{0.6}
\def\boxH{0.3}
\def\dA{1.8}
\def\dB{1.3}
\def\dC{2.4}
\def\wLine{0.6}

\def\figW{18}
\def\figH{0.6}

\tikzset{
    every node/.style={font=\footnotesize},
    every label/.style={font=\footnotesize}
}

        \draw[thick] (0,0) rectangle (\figW,\figH);
        \fill[fill={rgb,255:red,0;green,115;blue,189}] (\xA,0.5*\figH-0.5*\boxH) rectangle (\xA+\boxW,0.5*\figH+0.5*\boxH);
        \fill[fill={rgb,255:red,77;green,191;blue,237}] (\xA+\boxW,0.5*\figH-0.5*\boxH) rectangle (\xA+2*\boxW,0.5*\figH+0.5*\boxH);
        \node[right] at (\xA+2*\boxW,0.5*\figH) {$\theta$-component};  
        
        \fill[fill={rgb,255:red,0;green,115;blue,189}] (\xA+2*\boxW+\dA,0.5*\figH-0.5*\boxH) rectangle (\xA+3*\boxW+\dA,0.5*\figH+0.5*\boxH);
        \node[right] at (\xA+3*\boxW+\dA,0.5*\figH) {$\varphi = 0^\circ$};  
        
        \fill[fill={rgb,255:red,77;green,191;blue,237}] (\xA+2*\boxW+2*\dA,0.5*\figH-0.5*\boxH) rectangle (\xA+3*\boxW+2*\dA,0.5*\figH+0.5*\boxH);
        \node[right] at (\xA+3*\boxW+2*\dA,0.5*\figH) {$\varphi = 90^\circ$};
        \fill[fill={rgb,255:red,163;green,20;blue,46}] (\xA+\dist,0.5*\figH-0.5*\boxH) rectangle (\xA+\boxW+\dist,0.5*\figH+0.5*\boxH);
        \fill[fill={rgb,255:red,217;green,84;blue,26}] (\xA+\boxW+\dist,0.5*\figH-0.5*\boxH) rectangle (\xA+2*\boxW+\dist,0.5*\figH+0.5*\boxH);
        \node[right] at (\xA+2*\boxW+\dist,0.5*\figH) {$\varphi$-component};  
        
        \fill[fill={rgb,255:red,163;green,20;blue,46}] (\xA+2*\boxW+\dA+\dist,0.5*\figH-0.5*\boxH) rectangle (\xA+3*\boxW+\dA+\dist,0.5*\figH+0.5*\boxH);
        \node[right] at (\xA+3*\boxW+\dA+\dist,0.5*\figH) {$\varphi = 0^\circ$};  
        
        \fill[fill={rgb,255:red,217;green,84;blue,26}] (\xA+2*\boxW+2*\dA+\dist,0.5*\figH-0.5*\boxH) rectangle (\xA+3*\boxW+2*\dA+\dist,0.5*\figH+0.5*\boxH);
        \node[right] at (\xA+3*\boxW+2*\dA+\dist,0.5*\figH) {$\varphi = 90^\circ$}; 

        \draw[gray, very thick] (\xA+3*\boxW+2*\dA+\dist+\dB,0.5*\figH) -- (\xA+3*\boxW+2*\dA+\dist+\dB+\wLine,0.5*\figH);
        \node[right] at (\xA+3*\boxW+2*\dA+\dist+\dB+\wLine,0.5*\figH) {Measurement};
        \draw[gray, very thick, dash pattern=on 6pt off 4pt] (\xA+3*\boxW+2*\dA+\dist+\dB+\dC,0.5*\figH) -- (\xA+3*\boxW+2*\dA+\dist+\dB+\wLine+\dC,0.5*\figH);
        \node[right] at (\xA+3*\boxW+2*\dA+\dist+\dB+\wLine+\dC,0.5*\figH) {Simulation};




    \end{tikzpicture}

}
    
    \caption{Comparison of measured and simulated realized gain patterns of the fabricated four-port \ac{M3PA} at $f_c = \unit[28]{GHz}$ in the $xz$-plane ($\varphi = 0^\circ$), $yz$-plane ($\varphi = 90^\circ$), and $xy$-plane ($\theta = 90^\circ$) with respect to the coordinate system shown in \cref{fig:feeds}. (a)~-~(b)~Port~1. (c)~-~(d)~Port~2. (e)~-~(f)~Port~3. (g)~-~(h)~Port~4. (i)~Legend.}
    \label{fig:ff_elem}
\end{figure*}
\begin{figure}
    \centering
    \subfloat[\label{fig:ECC_elem:a}]{
    \includegraphics[]{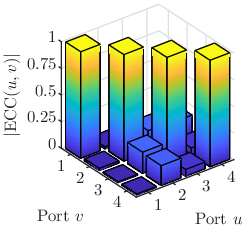}
    }
    \subfloat[\label{fig:ECC_elem:b}]{
    \includegraphics[]{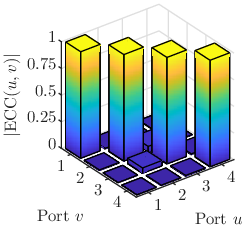}
    }
    \caption{\Acfp{ECC} of the \ac{M3PA} single antenna. (a)~\ac{ECC} calculated from measured radiation patterns. (b)~\ac{ECC} calculated from simulated radiation patterns.}
    \label{fig:ECC_elem}
\end{figure}

To verify the functionality of the proposed aperture radiator \ac{M3PA} concept, the single element has been built up and measured separately to the total array. A picture of the manufactured single element antenna as well as $4\times4$ array prototype is shown in \cref{fig:manufactured_prototypes}. 

The S-parameters of the antenna element have been measured using a four-port \ac{VNA}.
A comparison between the simulated and measured S-parameters of the single element \ac{M3PA} is shown in \cref{fig:spar_elem}. As can be seen from \cref{fig:spar_elem}, the ports of the antenna are sufficiently matched and stay below the limit of $|S_{uu}| \leq \unit[-10]{dB}$ within the target band n257 ($26.5\leq f/\mathrm{GHz}<29.5$). Further, highest measured coupling between the antenna ports is $|S_{uv,\mathrm{max}}|=\unit[-16]{dB}$ at the center frequency. However, most of the other coupling S-parameters stay well beyond $|S_{uv}|=\unit[-20]{dB}$. The reason for the increased coupling of ports 2 and 3 lies in the feed network. As can be seen from \cref{fig:antenna_concept}, ports 2 and 3 share the same branch-line coupler. As all of the modes belong to different orthogonal current groups, they theoretically do not couple. However, as the branch-line coupler is connected before the antenna, the finite coupling of the feed network directly translates to the S-parameters of the total antenna ports. 

For the verification of the far-field radiation performance, the single element antenna is installed and measured in a \ac{CATR}. The ports of the antenna are measured sequentially with all other ports terminated to $\unit[50]{\Omega}$. The 3D radiation patterns with amplitude and phase information are measured using a \ac{VNA}, utilizing the gain comparison method. The \ac{AUT} is installed on the positioner and rotated to obtain all measurement angles. It should be mentioned that there exists a blind spot due to the absorber-covered positioner on which the \ac{AUT} is mounted, in the range of $\theta\in180^\circ\pm30^\circ$.

The measured far-field pattern cuts of the single element antenna are shown in \cref{fig:ff_elem}. Especially the dipole ports 1 and 4 show very good agreement with the simulated antenna radiation patterns, while there are some minor deviations from the simulated patterns visible in the radiation patterns corresponding to ports 2 and 3. These deviations can potentially be attributed to the manufacturing process of the 3D-printed superstrate that is placed on top of the radiating aperture. 
First of all, the infill of the structure printed in the \ac{FDM} process directly affects the total dielectric constant of the superstrate. Any defects or gaps in the structure of the dielectric can decrease the total dielectric constant of the structure and lead to a mismatch between simulation and measurement. Furthermore, stair-casing of the \ac{FDM} manufacturing process can lead to micro-reflections (or internal reflections) of the transmitted waves and cause a distortion of the far-field radiation pattern. And finally, positioning inaccuracies of the dielectric on top of the aperture can lead to beam squint, causing the antenna patterns to shift towards one direction.

The \acfp{ECC} of the measured single element \ac{M3PA} are illustrated in \cref{fig:ECC_elem} and compared to the simulated results. As can be seen from \cref{fig:ECC_elem}, the \acp{ECC} of the measured antenna patterns agree generally well with the simulated \acp{ECC}. However, an increase of the \ac{ECC} of port 2 is visible which could be attributed to misalignment of the superstrate. Nevertheless, this verifies that the designed aperture-based \ac{M3PA} successfully excites far-fields corresponding to different orthogonal aperture modes. Furthermore, it validates that the individual antenna ports are sufficiently decoupled.

\subsection{Measurement and Validation of the Multi-Mode Multi-Port Antenna Array}

\begin{figure}
    \centering
    \subfloat[\label{fig:array_in_chamber:a}]{
    \input{figures/array_in_chamber}
    }
    \\
    \subfloat[\label{fig:array_in_chamber:b}]{
    \includegraphics[width=80mm]{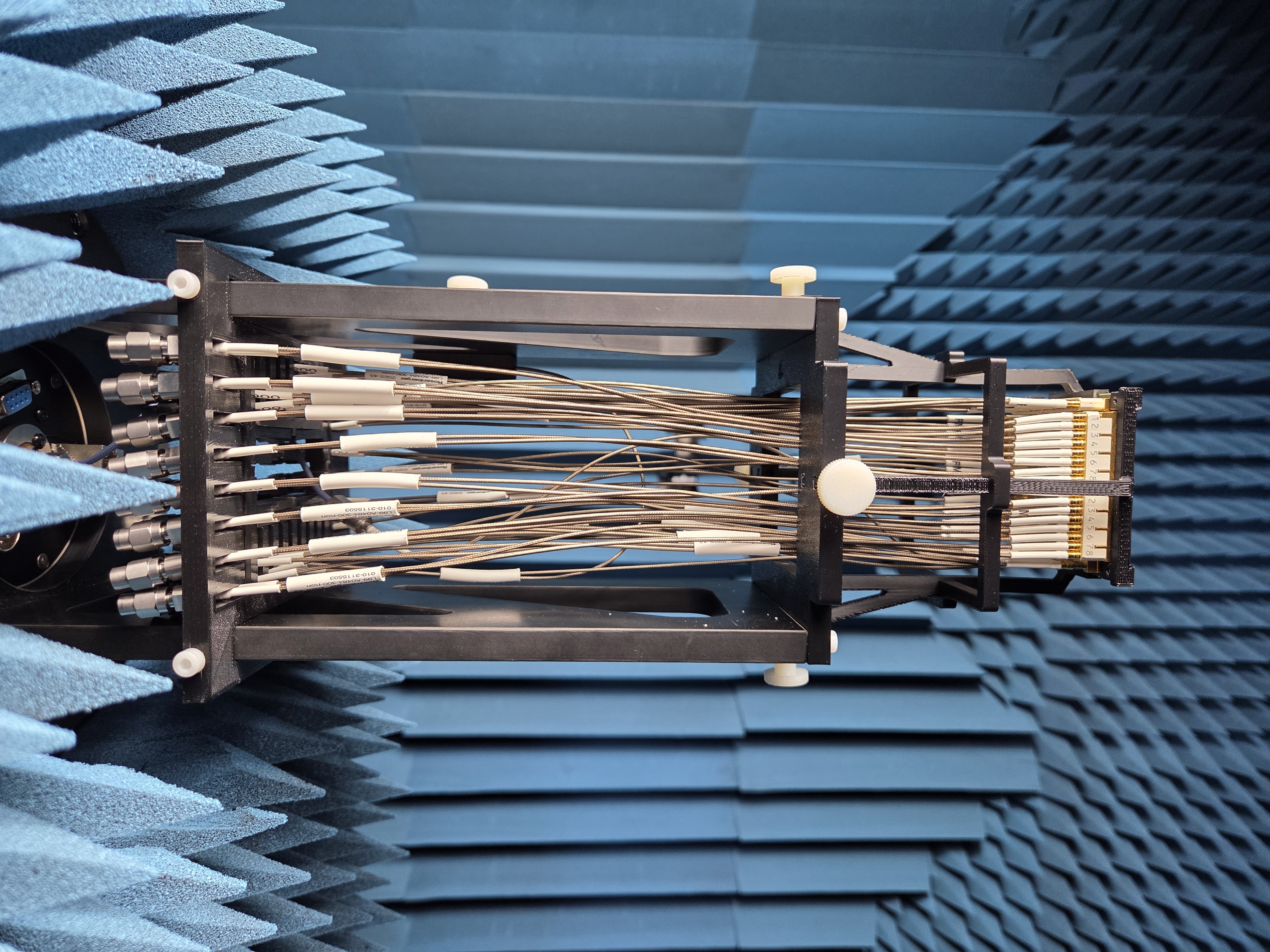}
    }
    \caption{Overview of the $4\times4$ \ac{M3PA} array installed inside the anechoic chamber. (a)~Front view. (b)~Side view with cables and terminations connected.}
    \label{fig:array_in_chamber}
\end{figure}

\begin{figure}
    \centering
    \subfloat[\label{fig:ff_array_theta}]{
    \includegraphics[]{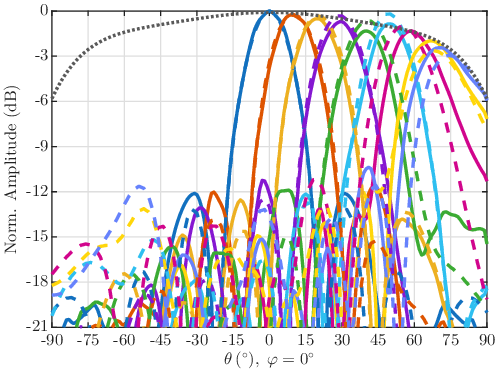}
    } \\
    \subfloat[\label{fig:ff_array_phi}]{
    \includegraphics[]{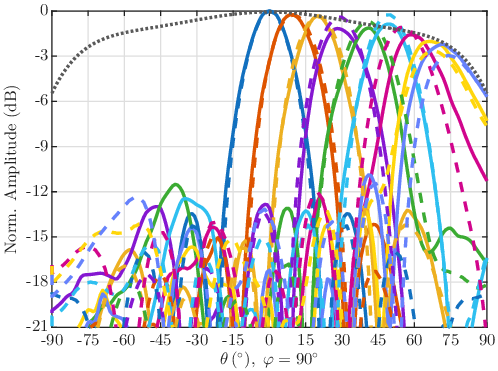}
    }\\
    \subfloat[\label{fig:ff_array_theta_leg}]{


\begin{tikzpicture}[every node/.style={font=\footnotesize}]
\newcommand{\figW}{8.6}
\newcommand{\figH}{1.4}

\newcommand{\xA}{0.2}
\newcommand{\lineW}{0.5}

\definecolor{colA}{RGB}{17, 113, 190}
\definecolor{colB}{RGB}{221, 84, 0}
\definecolor{colC}{RGB}{237, 177, 32}
\definecolor{colD}{RGB}{133, 22, 209}
\definecolor{colE}{RGB}{59, 170, 50}
\definecolor{colF}{RGB}{47, 190, 239}
\definecolor{colG}{RGB}{209, 4, 139}
\definecolor{colH}{RGB}{255, 214, 10}
\definecolor{colI}{RGB}{101, 130, 253}

\def\colorList{{"colA", "colB", "colC", "colD", "colE", "colF", "colG", "colH", "colI"}}
\definecolor{envGray}{rgb}{0.35, 0.35, 0.35}

\draw[thick] (0,0) rectangle (\figW,\figH);

    \newcounter{counter}
    \setcounter{counter}{0} 
    \foreach \x in{0,2,...,4} {
    \foreach \y in {1.1, 0.7, 0.3} {
        \stepcounter{counter}
        \pgfmathsetmacro{\currentCol}{\colorList[\the\value{counter}-1]}
        \pgfmathtruncatemacro {\currentAng}{(\the\value{counter}-1)*10}
        \draw[\currentCol, line width=1.5pt] (\xA+\x, \y) -- (\xA+\x+\lineW, \y) node[right,black] {$\theta = \currentAng^\circ$};
    }
    }

    \draw[gray, line width=1.5pt] (\xA + 6, 1.1) -- (\xA+\lineW + 6, 1.1) node[right,black] {Measurement};
    \draw[gray, line width=1.5pt,dashed] (\xA + 6, 0.7) -- (\xA+\lineW + 6, 0.7) node[right,black] {Simulation};
    \draw[envGray, line width=1.5pt,dotted] (\xA + 6, 0.3) -- (\xA+\lineW + 6, 0.3) node[right,black] {Envelope};


\end{tikzpicture}

    }
    \caption{Steered beams of the \ac{M3PA} array for different steering angles and polarizations at the target frequency of $f_c = \unit[28]{GHz}$. (a)~Norm. amplitude of the steered horizontally-polarized beams in $\theta$-direction for $\varphi=0^\circ$ ($xz$-plane). (b)~Norm. amplitude of the steered vertically-polarized beams in $\theta$-direction for $\varphi=90^\circ$ ($yz$-plane). (c)~Legend.
    }    
    \label{fig:steered_beams}
\end{figure}
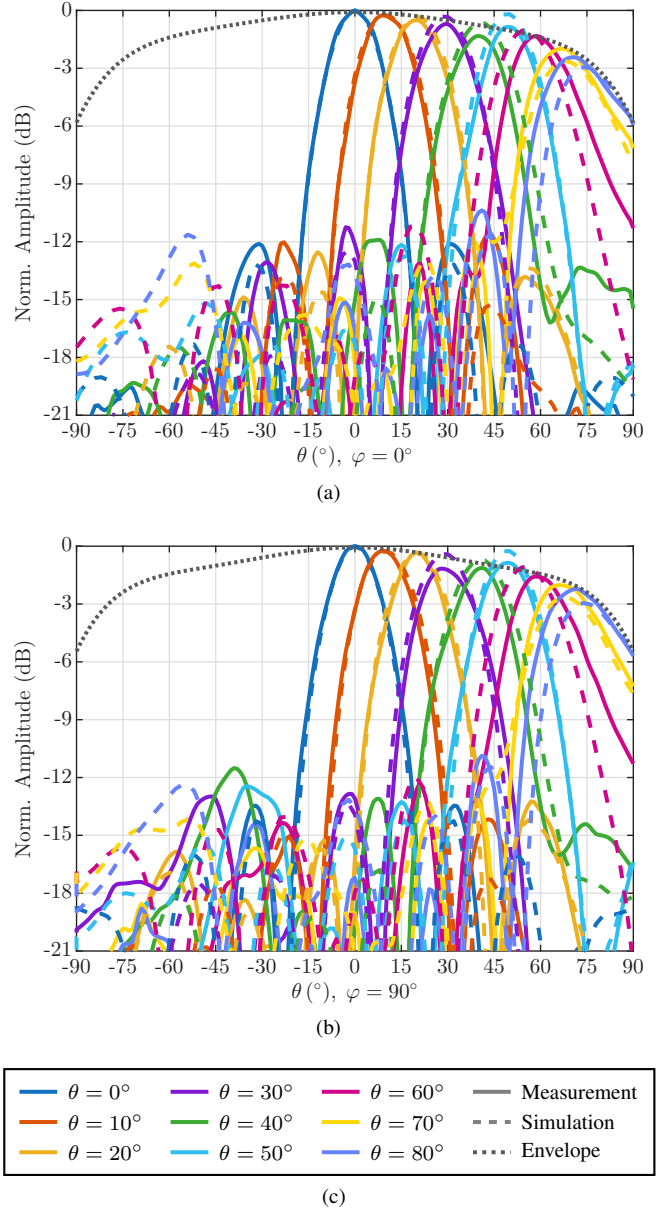

To evaluate the performance of the $4 \times 4$ \ac{M3PA} array, the array is measured within the \ac{CATR}. The $4\times4$ array installed on the positioner inside the \ac{CATR} is shown in \cref{fig:array_in_chamber}. The antenna ports of the array are measured sequentially to obtain the \acp{EEP} of all antenna ports. All unused ports are terminated with $\unit[50]{\Omega}$ terminations as shown in \cref{fig:array_in_chamber}~(b). As the \acp{EEP} are measured, all coupling effects are included within the measured patterns. To determine the total array pattern, the measured \acp{EEP} are superimposed using
\begin{equation}
    \vec{E}_\mathrm{Array} = \sum\limits_{n=1}^{N} a_n \vec{E}_{\mathrm{EEP},n}.
\end{equation}
Here, $a_n$ are the complex weighting coefficients of the individual antenna ports which are determined based on the algorithm shown in \cref{sec:bf_algorithm}. This method is in line with well-established concepts for the synthesis of total array patterns, compare \cite{Kelley1993} and is frequently used to determine the radiation characteristics of complete arrays \cite{Kasemodel2013,Yetisir2016}. 

\cref{fig:steered_beams} shows the steered total array far-field patterns for both horizontal and vertical polarization based on the measured \acp{EEP} of the $4\times4$ antenna array compared to the simulated results. For the synthesis of the array beams, the lower right corner of the array has been measured (ports $1 - 16$) and the far-fields were mirrored to obtain the \acp{EEP} of the full array. As visible from \cref{fig:steered_beams}, the measured steered beams match the simulated results well for the main lobe steered towards different target directions. The measured results confirm that the designed \ac{mmwave} \ac{M3PA} array is capable of achieving wide-angle steering performance up to a maximum angle of $\pm77^\circ$ for both the horizontal and vertical polarizations w.r.t. a maximum scan loss of $L_\mathrm{scan} = \unit[3]{dB}$. 
The array has a measured efficiency of $54.6\%$ for the vertically-polarized beam steered towards $\theta_0=40^\circ,\varphi_0=90^\circ$ compared to the simulated efficiency of $63.5\%$. The measured realized gain is $G_\text{real,meas.}=\unit[16.58]{dBi}$ in broadside direction compared to the simulated real. gain of $G_\text{real,sim.}=\unit[17.7]{dBi}$. Hence, the experimental results of the prototype $4\times4$ array fulfill the predefined project specifications. The decreased efficiency is attributed to the length of the breakout traces, routing the input signals from the connectors to the antenna elements. These traces cause a loss of approximately $L_\mathrm{Traces}=\unit[1]{dB}$, reducing the overall efficiency from $80\%$ at the feed network input to $63.5\%$ at the connector inputs. As mentioned earlier, access to all 64 ports is only required for this academic prototype. In an industrialized array implementation, these additional input losses would vanish, as each element port would be driven directly by a \ac{BFIC}. It is visible from \cref{fig:steered_beams}, that the wide-angle scanning performance is achieved with no visible grating lobes. 

\section{Conclusion}
This work has presented a methodology for extending the \acf{TCM} to aperture radiators and applying it to the design of aperture-based \acfp{M3PA}. The developed theory enables the realization of aperture antenna elements with multiple decoupled ports and radiation patterns, providing additional degrees of freedom for array beamforming and pattern synthesis. By exploiting broadside and off-broadside radiation modes, \acp{M3PA} offer significant potential to improve array performance at large steering angles.

To utilize these additional degrees of freedom, an iterative beamforming optimization algorithm has been developed to determine the excitation weights of the \ac{M3PA} array elements. This capability allows the adaptive placement of far-field nulls towards grating lobe directions and enables array designs with inter-element spacings exceeding the conventional half-wavelength limit.

The proposed concepts have been validated through the design of an aperture-based \ac{M3PA} element and its integration into a fully \ac{PCB}-implemented $4\times4$ antenna array. Measurements of the fabricated prototype confirm the proposed functionality, demonstrating wide-angle beam steering up to $\pm77^\circ$ with a scan loss of $L_\mathrm{scan}=\unit[3]{dB}$. These results highlight the potential of aperture-based \acp{M3PA} for wide-angle steering antenna arrays.

\appendix
\section*{Influence of the Superstrate on Beam Steering}
\label{appendix}
To evaluate the influence of the superstrate on the steering performance, the array element is simulated with the superstrate placed on the aperture and without it and arranged in a $4\times4$ array configuration. The inter-element distance is set to $d = \unit[6.5]{mm} \approx 0.6\lambda_0$.
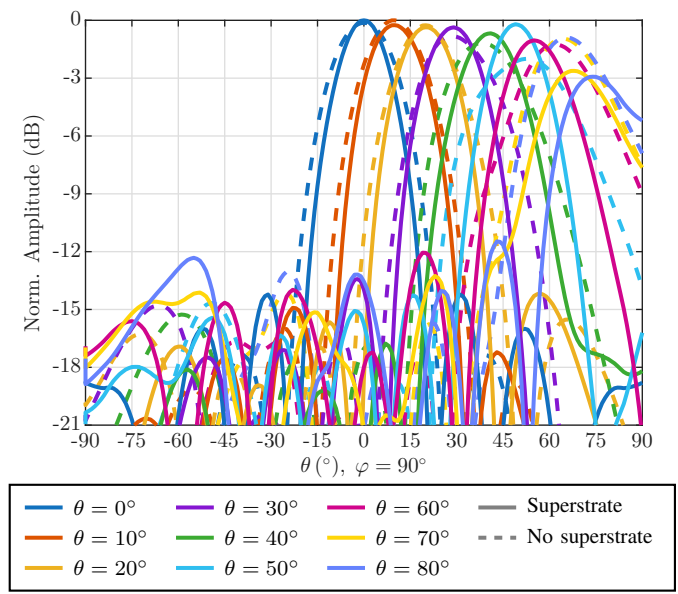
\begin{figure}[b]
    \centering
    \includegraphics[]{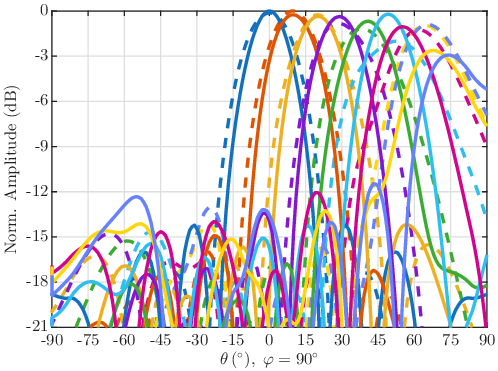}
    \vspace{1pt}


\begin{tikzpicture}[every node/.style={font=\footnotesize}]
\newcommand{\figW}{8.8}
\newcommand{\figH}{1.4}

\newcommand{\xA}{0.2}
\newcommand{\lineW}{0.5}

\definecolor{colA}{RGB}{17, 113, 190}
\definecolor{colB}{RGB}{221, 84, 0}
\definecolor{colC}{RGB}{237, 177, 32}
\definecolor{colD}{RGB}{133, 22, 209}
\definecolor{colE}{RGB}{59, 170, 50}
\definecolor{colF}{RGB}{47, 190, 239}
\definecolor{colG}{RGB}{209, 4, 139}
\definecolor{colH}{RGB}{255, 214, 10}
\definecolor{colI}{RGB}{101, 130, 253}

\def\colorList{{"colA", "colB", "colC", "colD", "colE", "colF", "colG", "colH", "colI"}}
\definecolor{envGray}{rgb}{0.35, 0.35, 0.35}

\draw[thick] (0,0) rectangle (\figW,\figH);

    \setcounter{counter}{0} 
    \foreach \x in{0,2,...,4} {
    \foreach \y in {1.1, 0.7, 0.3} {
        \stepcounter{counter}
        \pgfmathsetmacro{\currentCol}{\colorList[\the\value{counter}-1]}
        \pgfmathtruncatemacro {\currentAng}{(\the\value{counter}-1)*10}
        \draw[\currentCol, line width=1.5pt] (\xA+\x, \y) -- (\xA+\x+\lineW, \y) node[right,black] {$\theta = \currentAng^\circ$};
    }
    }

    \draw[gray, line width=1.5pt] (\xA + 6, 1.1) -- (\xA+\lineW + 6, 1.1) node[right,black] {Superstrate};
    \draw[gray, line width=1.5pt,dashed] (\xA + 6, 0.7) -- (\xA+\lineW + 6, 0.7) node[right,black] {No superstrate};


\end{tikzpicture}

    \caption{Comparison of the vertically-polarized steered beams of the simulated antenna array with and without superstrate for $\varphi=90^\circ$ ($yz$-plane).}
    \label{fig:compare_diel}
\end{figure}
\Cref{fig:compare_diel} shows a comparison between the simulated steered beams of the $4\times4$ array in this configuration. Due to the addition of the superstrate, the beam width of the steered beams is decreased while increasing the gain by about $\unit[1]{dB}$. Further, a slight increase in the sidelobe level is visible. However, in both cases the maximum achievable scanning range w.r.t. a scan loss of $L_\mathrm{scan}=\unit[3]{dB}$ remains equal with approximately $77^\circ$.




\bibliographystyle{IEEEtran.bst}
\bibliography{IEEEabrv.bib,references.bib}

\begin{IEEEbiography}[{\includegraphics[width=1in,height=1.25in,clip,keepaspectratio]{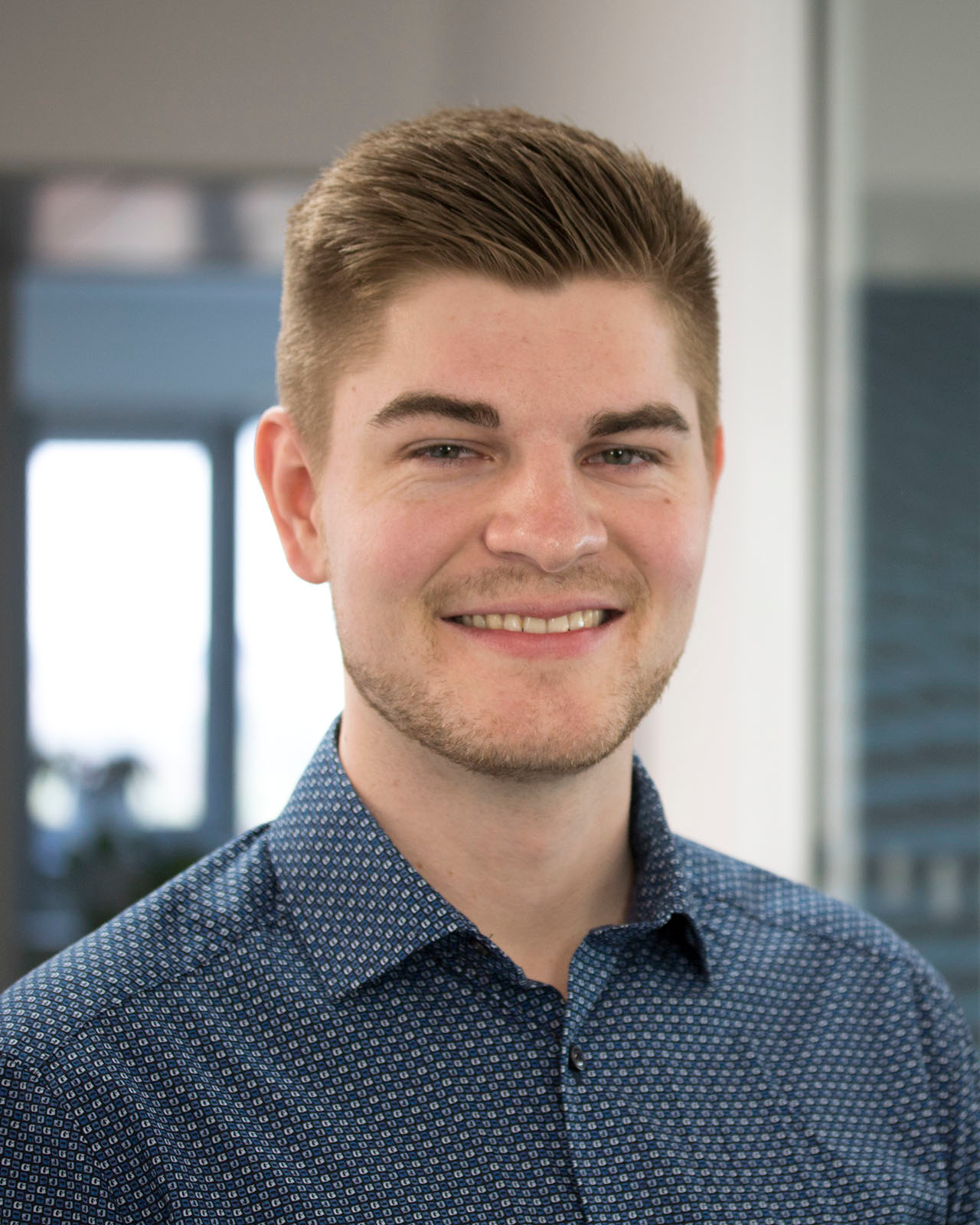}}]{Tim Hahn} 
(Graduate Student Member, IEEE) was born in Hannover, Germany, in 1997. He received the B.Sc. and M.Sc. degrees in electrical engineering from Leibniz University Hannover, Hannover in 2020 and 2021, respectively. He is currently a Research Assistant with the Institute of Microwave and Wireless Systems, Leibniz University Hannover where he is pursuing his Ph.D. in Electrical Engineering and Information Technology. His current research focuses on antenna array design based on characteristic modes, modal analysis of aperture radiators, and enhanced beamforming techniques using multi-mode multi-port antennas. Some of his further fields of interest include millimeter wave antenna integration for mobile terminals, computational electromagnetics, and RF circuit design.
\end{IEEEbiography}

\begin{IEEEbiography}[{\includegraphics[width=1in,height=1.25in,clip,keepaspectratio]{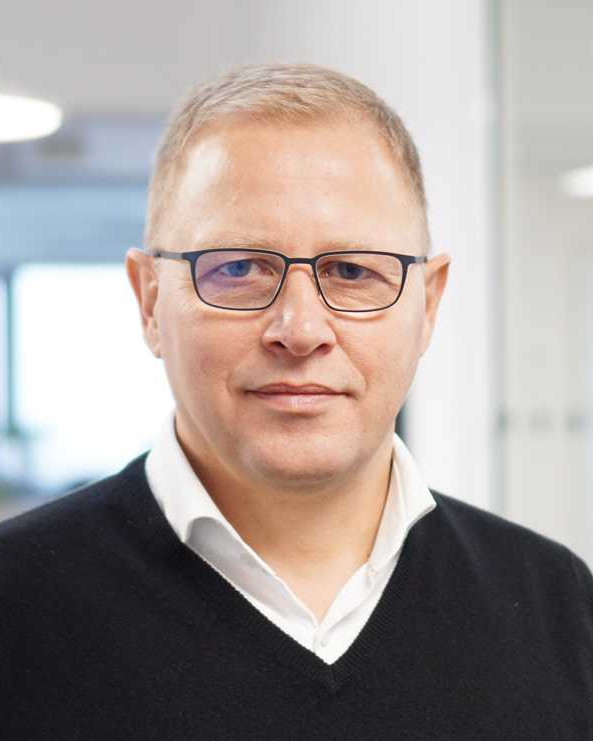}}]{Dirk Manteuffel} 
(Member, IEEE) was born in Issum, Germany, in 1970. He received the Dipl.-Ing. and Dr.-Ing. degrees in electrical engineering from the University of Duisburg–Essen, Duisburg, Germany, in 1998 and 2002, respectively. From	1998 to 2009, he was with IMST, Kamp-Lintfort, Germany. As a Project Manager, he was responsible for industrial antenna development and advanced projects in the field of antennas and electromagnetic (EM) modeling. From 2009 to 2016, he was a Full Professor of wireless communications at Christian-Albrechts-University, Kiel, Germany. Since June 2016, he has been a Full Professor and the Executive Director of the Institute of Microwave and Wireless Systems, Leibniz University Hannover, Hannover, Germany. His research interests include electromagnetics, antenna integration and EM modeling for mobile communications and biomedical applications. Dr. Manteuffel was a director of the European Association on Antennas and Propagation from 2012 to 2015. Heserved on the Administrative Committee (AdCom) of IEEE Antennas and Propagation Society from 2013 to 2015 and as an Associate Editor of the IEEE Transactions on Antennas and Propagation from 2014 to 2022. Since 2009 he has been an appointed member of the committee "Antennas" of the German VDI-ITG.
\end{IEEEbiography}


\begin{acronym}
    \acro{TCM}[TCM]{theory of characteristic modes}
    \acro{CM}[CM]{characteristic mode}
    \acro{CMA}[CMA]{characteristic mode analysis}
	\acro{EFIE}[EFIE]{electric-field integral equation}
	\acro{FDTD}[FDTD]{finite differences time domain}
	\acro{M3PA}[M³PA]{multi-mode multi-port antenna}
	\acro{MoM}[MoM]{method of moments}
	\acro{PEC}[PEC]{perfect electric conductor}
	\acro{PMC}[PMC]{perfect magnetic conductor}
    \acro{EM}[EM]{electromagnetic}
	\acro{RWG}[RWG]{Rao-Wilton-Glisson}
	\acro{mmwave}[mmWave]{millimeter wave}
    \acro{DUT}[DUT]{device under test}
    \acro{ECC}[ECC]{envelope correlation coefficient}
    \acro{IRP}[irrep]{irreducible representation}
    \acro{PCB}[PCB]{printed circuit board}
    \acro{EEP}[EEP]{embedded element pattern}
    \acro{AEEP}[AEEP]{average embedded element pattern}
    \acro{CATR}[CATR]{compact antenna test range}
    \acro{HPBW}[HPBW]{half power beam width}
    \acro{FDM}[FDM]{fused deposition modeling}
    \acro{VNA}[VNA]{vector network analyzer}
    \acro{AUT}[AUT]{antenna under test}
    \acro{ISAC}[ISAC]{integrated sensing and communication}
    \acro{EBG}[EBG]{electromagnetic band gap}
    \acro{GEP}[GEP]{generalized eigenvalue problem}
    \acro{MIMO}[MIMO]{multiple-input multiple-output}
    \acro{ICC}[ICC]{current correlation coefficient}
    \acro{BFIC}[BFIC]{beamforming integrated circuit}
\end{acronym}

\end{document}